\documentclass[12pt,preprint]{aastex}
\begin{document}
\newcommand{\degree}{\ensuremath{^\circ}}

\title{Type II Quasars from the Sloan Digital Sky Survey: \\ V. Imaging host galaxies with the Hubble Space Telescope$^{1,2}$.}

\author{
Nadia L. Zakamska\altaffilmark{3,4,5}, 
Michael A. Strauss\altaffilmark{4}, 
Julian H. Krolik\altaffilmark{6},
Susan E. Ridgway\altaffilmark{6}, \\ 
Gary D. Schmidt\altaffilmark{7},
Paul S. Smith\altaffilmark{7},
Timothy M. Heckman\altaffilmark{6},
Donald P. Schneider\altaffilmark{8},\\
Lei Hao\altaffilmark{9},
J. Brinkmann\altaffilmark{10}
\altaffiltext{1}{This paper is dedicated to the memory of John Norris Bahcall, a pioneer in the study of quasar host galaxies.}
\altaffiltext{2}{The observations reported here were obtained with the NASA/ESA Hubble Space Telescope and at the MMT Observatory, a facility operated jointly by the Smithsonian Institution and the University of Arizona. Public Access time is available at the MMT Observatory through an agreement with the National Science Foundation.}
\altaffiltext{3}{Institute for Advanced Study, Einstein Dr., Princeton, NJ 08540}
\altaffiltext{4}{Princeton University Observatory, Princeton, NJ 08544}
\altaffiltext{5}{Spitzer fellow}
\altaffiltext{6}{Department of Physics and Astronomy, Johns Hopkins University, 3400 North Charles Street, Baltimore, MD 21218-2686}
\altaffiltext{7}{Steward Observatory, The University of Arizona, 933 North Cherry Avenue, Tucson, AZ 85721}
\altaffiltext{8}{Department of Astronomy and Astrophysics, 525 Davey Laboratory, Pennsylvania State University, University Park, PA 16802}
\altaffiltext{9}{108 Space Sciences Building, Cornell University, Ithaca, NY 14853}
\altaffiltext{10}{Apache Point Observatory, P.O. Box 59, Sunspot, NM 88349}
}

\begin{abstract}
Type II quasars are luminous Active Galactic Nuclei whose centers are obscured by large amounts of gas and dust. In this paper we present 3-band HST images of nine type II quasars with redshifts $0.2<z<0.4$ selected from the Sloan Digital Sky Survey based on their emission line properties. The intrinsic luminosities of these AGNs are estimated to be $-24>M_B>-26$, but optical obscuration allows their host galaxies to be studied unencumbered by bright nuclei. Each object has been imaged in three continuum filters (`UV', `blue' and `yellow') placed between the strong emission lines. The spectacular, high quality images reveal a wealth of details about the structure of the host galaxies and their environments. Six of the nine galaxies in the sample are ellipticals with de Vaucouleurs light profiles, one object has a well-defined disk component and the remaining two have marginal disks. Stellar populations of type II quasar hosts are more luminous (by a median of 0.3$-$0.7 mag, depending on the wavelength) and bluer (by about 0.4 mag) than are $M_*$ galaxies at the same redshift. When smooth fits to stellar light are subtracted from the images, we find both positive and negative residuals that become more prominent toward shorter wavelengths. We argue that the negative residuals are due to kpc-scale dust obscuration, while most positive residuals are due to the light from the nucleus scattered off interstellar material in the host galaxy. Scattered light makes a significant contribution to the broad band continuum emission and can be the dominant component of the extended emission in the UV in extreme cases.
\end{abstract}


\section{Introduction}

One of the key long-term questions in quasar studies is to identify just which properties of the galactic host (overall morphology? presence of bars? close neighbors?  star-formation history? gas content?) are associated with activity in its nucleus. Observations of quasar hosts are made difficult by the small angular scales involved (1\arcsec$\simeq$ 4.5 kpc for $z=0.3$). Even for quasars with modest redshifts ($z\le 0.8$), optical imaging from the ground requires excellent seeing conditions or the use of adaptive optics (\citealt{ronn96, perc01, jahn03, ornd03, jahn04}; see \citealt{bahc97} for earlier references). Imaging of quasar hosts can therefore greatly benefit from the superior angular resolution in space and was one of the major science drivers of the Hubble Space Telescope (HST; \citealt{bahc97, boyc98, boyc99, mclu99, hami02, dunl03, paga03, floy04}). An additional difficulty present in ground-based and space-based observations is posed by the bright source in the center of the host galaxy -- the quasar itself. In many cases the light from the quasar completely overwhelms the host galaxy, preventing its detection altogether (e.g., \citealt{hami02}). Indeed, the original definition of a quasar required a point-like morphology.

In the context of lower luminosity Active Galactic Nuclei (Seyfert galaxies), much of the work on host galaxies has focused on ``type 2" objects, Seyfert nuclei deeply obscured along our line of sight by matter close to the nucleus \citep{heck97, malk98, mart99, stor01, gonz02, kauf03}. There is strong evidence that type 2 Seyfert galaxies would be classified as ordinary, ``type 1" objects if viewed from a different direction, where the obscuration is negligible (for reviews, see \citealt{anto93, krol99}). The hosts of type 2 Seyferts should then form a representative sample for the hosts of all Seyfert galaxies. The advantage of using type 2 Seyferts is that images of their hosts may be studied unencumbered by bright nuclei. Subtracting the light from the active nucleus is especially problematic at quasar luminosities, and a significant advantage could be gained by studying the hosts of obscured quasars. Investigation of obscured quasar hosts is the subject of the present paper.

Although the direct line of sight to the nucleus is deeply obscured, nuclear light can contribute to the extended emission by way of scattering in the host galaxy \citep{anto85}. The scattering efficiency (the ratio of the apparent luminosity due to scattering to the intrinsic luminosity) is very poorly known and certainly varies by a large factor from object to object, but some observations suggest that it can be as large as a few per cent \citep{zaka05}. If the nuclear luminosity is $\sim 100\times$ the host luminosity, a scattering efficiency of 1\% doubles the apparent luminosity of the host. While such contrast between the nuclear and the host luminosities is rare in the $V$ band \citep{bahc97}, it might be common in the blue and near-UV, so we expect a significant contribution of the scattered light in these bands. Aided by polarimetric data and detailed morphological information, we attempt to separate the scattered light contribution from the stellar component of the host galaxy, but there are often severe limits on how well this can be done. 

In this paper we present results of the imaging of nine type II (obscured) quasars using the Advanced Camera for Surveys (ACS) aboard the HST, in combination with polarimetric data from ground-based observations. In Section \ref{sec_sample} we describe the sample selection, observations and data reduction. We describe our tools for image analysis in Section \ref{sec_image_analysis}. Individual objects are described in Section \ref{sec_ind}. We discuss our results in Section \ref{sec_discussion} and conclude in Section \ref{sec_conclusions}. Throughout this paper, we use a cosmology with $h=0.7$, $\Omega_m=0.3$ and $\Omega_{\Lambda}=0.7$. Objects are identified as SDSS Jhhmmss.ss$+$ddmmss.s in Table \ref{tab:ids} (e.g., SDSS~J103951.49+643004.2) and as SDSS Jhhmm$+$ddmm elsewhere. Spectral features are identified using air wavelengths in angstroms (e.g., [OIII]5007).

\section{Sample selection and observations}
\label{sec_sample}

\subsection{Sample selection}

The selection and optical properties of the parent sample of 291 type II Active Galactic Nuclei (AGNs) candidates were described by \citet{zaka03}, hereafter Paper I. Briefly, these objects were identified in the spectroscopic database of the Sloan Digital Sky Survey (SDSS; \citealt{york00, stou02, abaz04}) as sources at redshifts $0.3<z<0.8$ showing narrow emission lines with high-ionization line ratios characteristic of non-stellar ionizing continua. We used the [OIII]5007 emission line as a proxy for the intrinsic luminosity and estimated that those objects with [OIII]5007 emission line luminosities in excess of $3\times 10^8 L_{\odot}$ have intrinsic luminosities $M_B<-23$. This luminous subsample consists of about 150 objects which we classified as type II quasars (rather than Seyfert 2 galaxies). High bolometric luminosities (well in excess of the conventional quasar-defining value of $10^{45}$ erg sec$^{-1}$) of many of these objects and high densities of neutral gas along the line of sight were confirmed by infrared and X-ray observations \citep{zaka04, vign04, ptak06}. 

At lower redshifts ($0<z<0.33$), type II AGNs from the SDSS have been studied extensively by \citet{kauf03} and \citet{hao05a, hao05b}. These samples were selected based on the emission line properties from the complete sample of SDSS galaxies \citep{stra02}, but only a small number of objects can be classified as type II quasars based on the aforementioned [OIII]5007 luminosity criterion, with the majority of the objects being Seyfert 2 galaxies. 

We then restricted ourselves only to those of the type II AGNs from the samples by \citet{zaka03} (Paper I) and by \citet{hao05a} which have $L$([OIII]5007)$>10^9L_{\odot}$ (i.e., well above the quasar luminosity criterion) and which were radio-quiet \citep{zaka04}. Since the equivalent widths of the emission lines in type II quasars can be very high (above $\sim 1000$\AA, Paper I), these lines can make a significant contribution to the broad-band flux. To avoid contamination from the strongest lines ([OII]3727, [OIII]4959,5007, H$\beta$, H$\alpha$) to the HST images, we selected out objects from two narrow redshift ranges so that three ACS broad-band filters could be placed between these lines. We selected three objects at $z\simeq 0.25$ from the sample by \citet{hao05a} and six objects at $z\simeq 0.4$ from the sample in Paper I. SDSS spectra of these objects are shown in Figure \ref{pic_spectra} together with the wavelength coverage of the filters that we used in our ACS observations. Using the [OIII]5007 line luminosities, we estimate that the intrinsic luminosities of the objects described in this paper are about $-24>M_B>-26$. 

Throughout this paper, we refer to the filters used in our observations as `yellow', `blue' and `UV', in order of decreasing wavelength. The `yellow' band has an effective rest-frame wavelength of 5700$-$5800\AA\ for all objects in our sample, i.e., somewhat redder than but close to the rest-frame $V$ band, and the `blue' and `UV' bands are longward and shortward of [OII]3727, respectively.

\subsection{HST observations and data reduction}

\begin{figure}
\epsscale{0.85}
\plotone{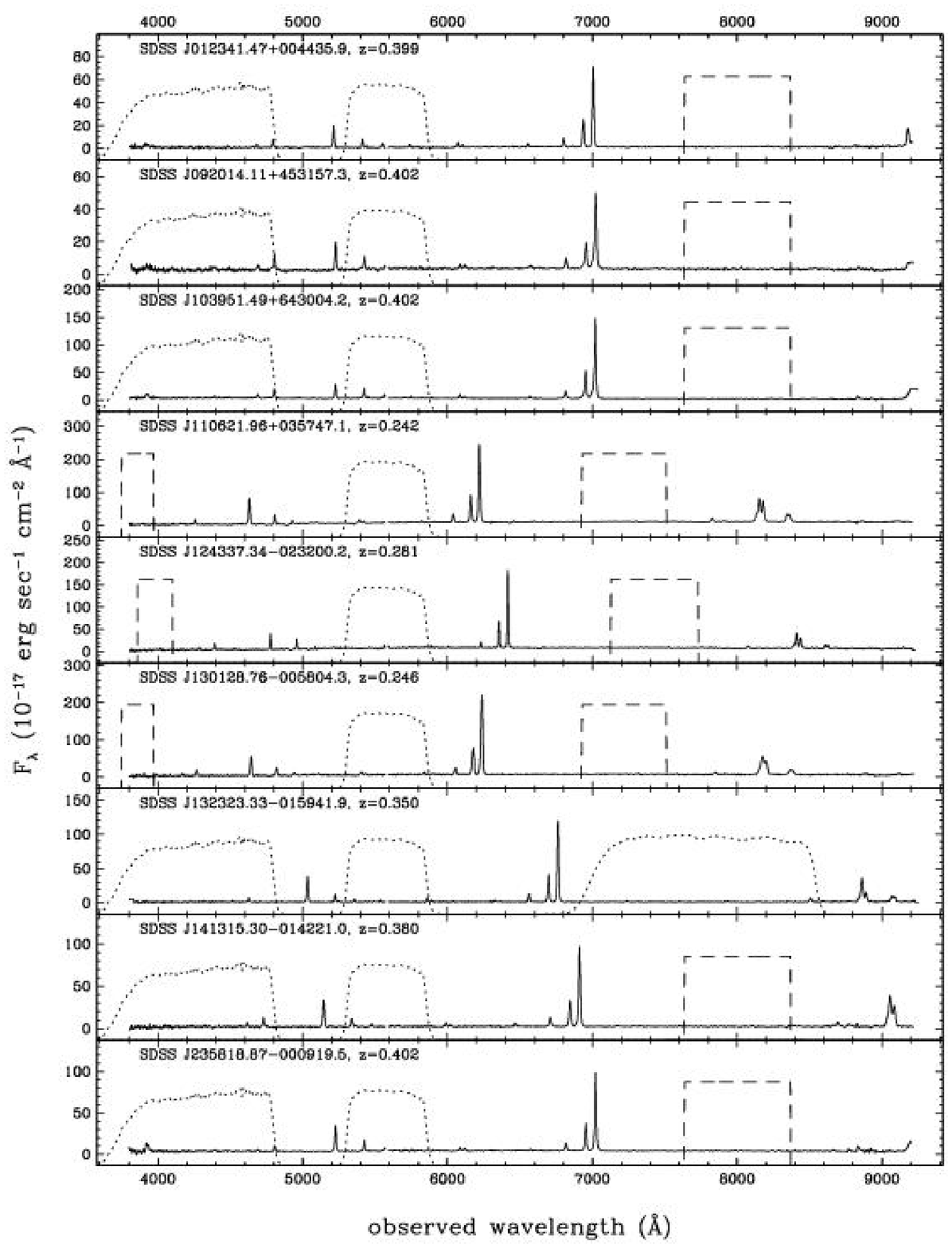}
\caption{SDSS spectra of the nine objects that are the subject of this paper. The resolution of the original spectra is $\lambda/\Delta \lambda \approx 2000$; the spectra have been smoothed by 5 pixels for display purposes. Overlaid on the spectra are transmittance curves of the regular ACS filters (dotted lines) and ramp ACS filters (dashed lines, schematic representation) used in our imaging program, as summarized in Table \ref{tab:ids}. The filters were chosen to avoid strong emission lines. Transmittance curves and widths of ramp filters were taken from the ACS instrument website, http://acs.pha.jhu.edu/instrument/.}
\label{pic_spectra}
\end{figure}

All observations pertaining to our program were obtained in the period July 2003 -- May 2004. Each object was imaged with the Wide Field Channel (image scale $0.049\arcsec$ pixel$^{-1}$) of the ACS in three filters, as summarized in Table \ref{tab:ids} and illustrated in Figure \ref{pic_spectra}. For each filter, four exposures were taken, each lasting 540-610 sec, in a four-point dither pattern, or overall observing time of one orbit per object per filter. Data reduction consisted of rejecting cosmic rays from each of the four flat-field-corrected exposures, applying a geometric distortion correction and then combining the four images. All data reduction was performed using the {\sl MultiDrizzle} routine \citep{koek02} available as part of the HST archive pipeline. 

Using the standard ACS photometric calibration described in the \citet{acs05}, we converted the units of the reduced data (counts/sec) into spectral flux densities $F_{\nu}$ and AB magnitudes $m_{AB}=-2.5 \log F_{\nu}-48.60$, where $F_{\nu}$ is in units of erg sec$^{-1}$ cm$^{-2}$ Hz$^{-1}$. The absolute AB magnitudes were calculated at rest-frame effective wavelengths $\lambda_{obs}/(1+z)$ as $M_{AB}=m_{AB}-5\log(D_L/10\mbox{ pc})+2.5\log(1+z)$, where $D_L$ is the luminosity distance and $\lambda_{obs}$ are the effective wavelengths in the observer's frame. The values $\lambda_{obs}/(1+z)$ are listed in Table \ref{tab:ids}. 

\begin{figure}
\epsscale{0.9}
\plotone{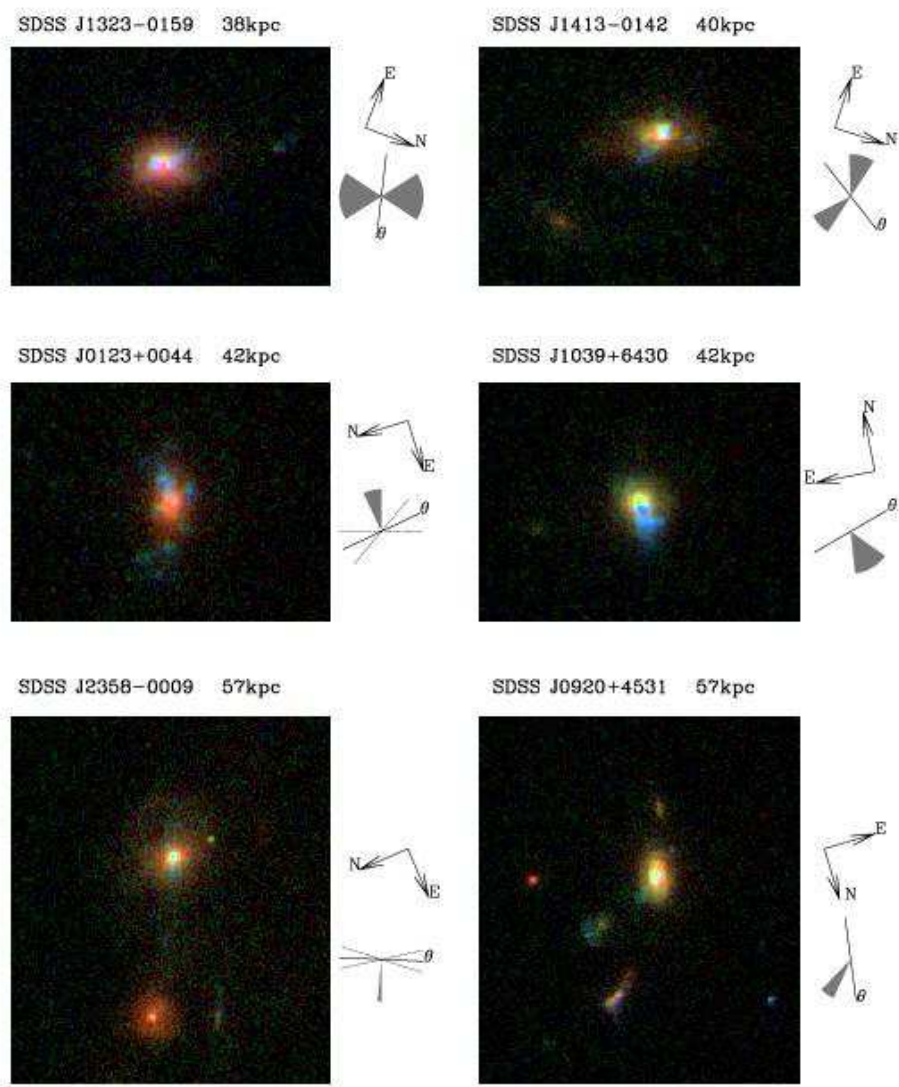}
\caption{Color-composite images of type II quasars produced using the algorithm by \citet{lupt04}. In all images except SDSS~1323$-$0159 (top left) the same parameters of the color-combining code were used (group 1 objects, see Table \ref{tab:ids}), so the colors of the objects can be directly visually compared. Object identification and the total horizontal size of the images are given above each frame. A cartoon of the presumed scattering regions and polarization position angle (in image coordinates), as well as the orientation on the sky are shown to the right of each image. The polarization position angle $\theta$ is marked with a solid black line, and grey lines mark the $1\sigma$ confidence limits if the uncertainty in the angle is significant. The brightest UV regions tentatively identified as scattered light are schematically shown as grey wedges.}
\label{pic_rgb1}
\end{figure}

\begin{figure}
\epsscale{0.9}
\plotone{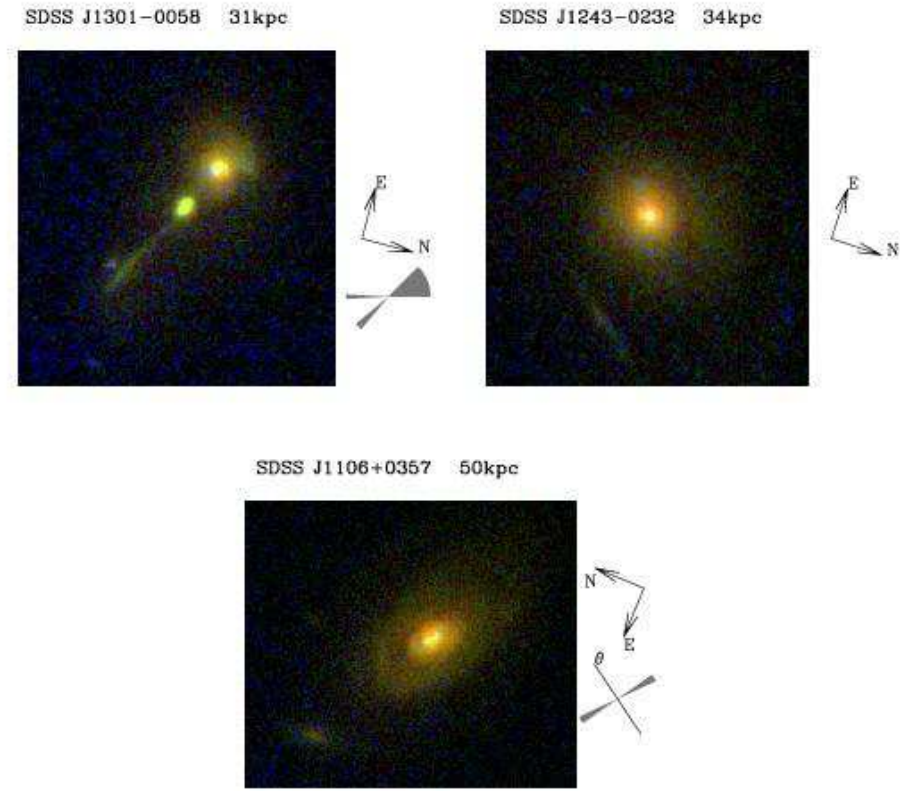}
\caption{Color-composite images of type II quasars produced using the algorithm by \citet{lupt04}. In all three images, the same parameters of the color-combining code were used (group 2 objects, see Table \ref{tab:ids}), although the positions of the ramp filters used for SDSS~J1243$-$0232 were slightly different from those used for the other two objects (Figure \ref{pic_spectra} and Table \ref{tab:ids}). The UV filter used for the three objects in this image (FR459M) is about an order of magnitude less sensitive than F435W used in Figure \ref{pic_rgb1}, so the scattering regions are harder to identify, and shot noise can be seen in the form of blue spots on top of a dark background. Images are labeled as in Figure \ref{pic_rgb1}. We do not have polarimetric measurements of SDSS~J1243$-$0232 and SDSS~J1301$-$0058.}
\label{pic_rgb2}
\end{figure}

The images taken in three bands were rebinned to the common grid using point sources in the field and combined to produce a color-composite image using the {\it asinh} stretching and color-combining routine by \citet{lupt04}. The resulting images are presented in Figures \ref{pic_rgb1}-\ref{pic_rgb2}. We used identical image combining parameters for the sources with similar redshifts that used similar or identical filters (one group of five objects and another group of three objects, Table \ref{tab:ids}). As a result, images from the same group can be directly compared with one another. For example, the yellow color of SDSS~J0139+6430 compared to the red color of SDSS~J0123+0044 reflects a difference in their stellar populations and is not an image processing artifact or the effects of K-corrections. The images in Figure \ref{pic_rgb2} (group 2 objects) have a rather different color scheme from those in Figure \ref{pic_rgb1}. This difference is due to using the ramp filter FR459M as the UV filter for the group 2 objects; this filter is significantly narrower than all other filters in our program (Figure \ref{pic_spectra}), resulting in about a factor of 10 reduction in sensitivity. Finally, SDSS~J1323$-$0159 was not grouped with any other sources because it is the only object that uses F775W as its yellow filter. 

\subsection{Optical polarimetry and spectropolarimetry}
\label{sec_pol}

In Table \ref{tab:pol} we summarize all polarimetric and spectropolarimetric data available for the objects observed with HST. Out of nine objects, broad-band polarimetry is available for two objects and spectropolarimetric data are available for five objects. Of the latter, three objects (SDSS~J1039+6430, SDSS~J1323$-$0159, and SDSS~J1413$-$0142) were previously reported by \citet{zaka05} (hereafter Paper II), and the results of these observations are given in Table \ref{tab:pol}. In this section, we describe spectropolarimetric and polarimetric observations of four more objects.

Spectropolarimetry was obtained for two objects, SDSS~J0920+4531 and SDSS~J1106+0357, on December 19, 2004. We used the CCD Spectropolarimeter (SPOL, \citealt{schm92b}) at the 6.5m Multiple Mirror Telescope (MMT). All observations used a low-resolution grating providing spectral coverage of 4100-8200\AA. With seeing of more than 2\arcsec\ (full width at half maximum), an entrance slit of 1.5\arcsec\ width was used, resulting in a spectral resolution of $\sim$22\AA. Each object was observed for 6400 sec. The absolute flux calibration was performed using the original SDSS spectra (whose spectrophotometric calibration is accurate to 5\%, \citealt{abaz04}) rather than photometric standards because seeing was highly variable on the night of the observations. The instrumental polarization was verified during the runs to be $<0.1\%$ through observations of unpolarized standard stars taken from \citet{schm92a}, and the polarization due to intervening Galactic dust is $< 0.2\%$ and $< 0.5\%$ for SDSS~J0920+4531 and SDSS~J1106+0357, respectively (estimated using the Galactic reddening values by \citealt{schl98} and the maximum interstellar polarization per unit reddening by \citealt{serk75}). Other details of instrument configuration, calibration and data reduction are the same as described in Paper II. 

\begin{figure}
\epsscale{0.9}
\plotone{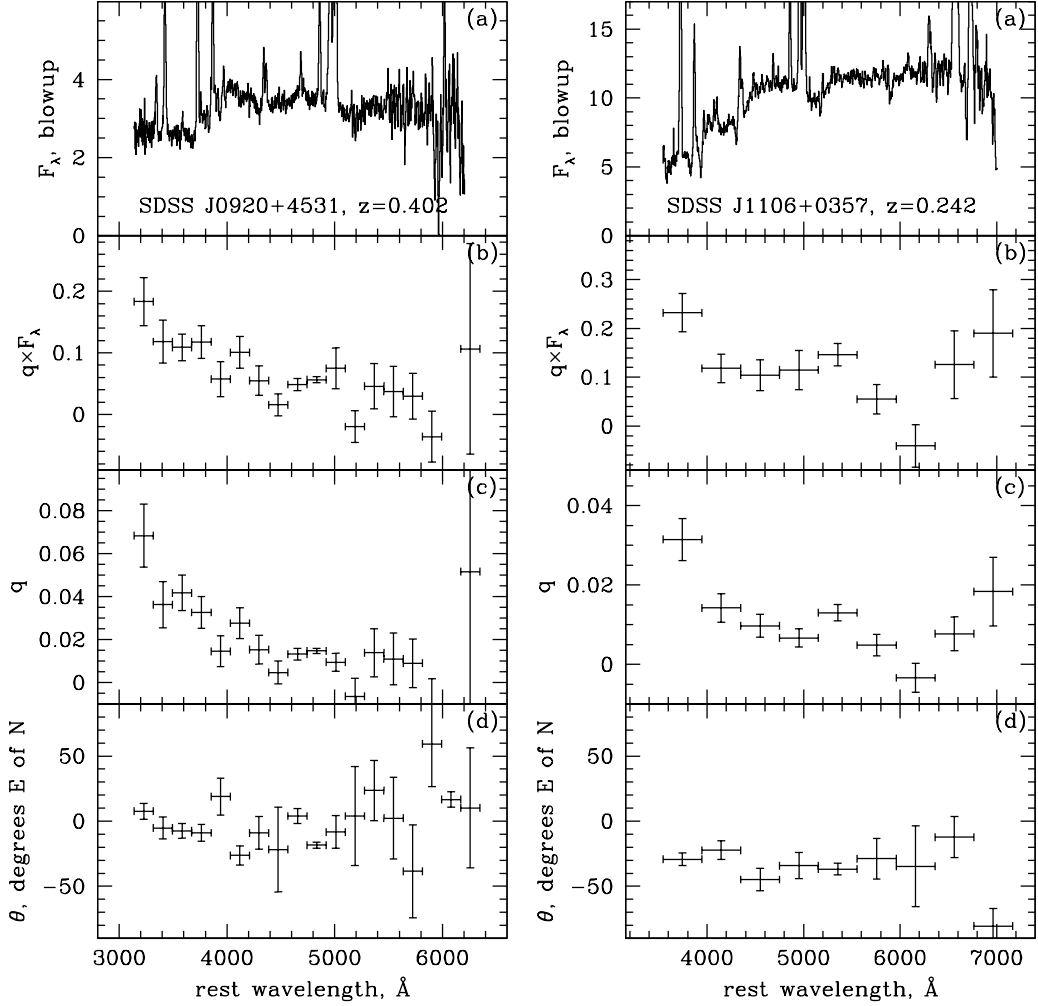}
\caption{Results of spectropolarimetry conducted using SPOL on MMT for SDSS~J0920+4531 (left) and SDSS~J1106+0357 (right). From top to bottom: (a) total optical flux density (focusing on the continuum); (b) polarized flux density; (c) polarization fraction; (d) polarization position angle in degrees east of north. Flux densities in panels (a) and (b) are in units of 10$^{-17}$ erg sec$^{-1}$ cm$^{-2}$ \AA$^{-1}$. All measurements are corrected for Galactic extinction and atmospheric absorption bands. Polarization measurements have been heavily binned in wavelength. The horizontal error bars reflect the bin size, whereas the vertical error bars are 1$\sigma$ uncertainties computed from the scatter of values within each bin.}
\label{pic_mmt}
\end{figure}

The results of the spectropolarimetric measurements are presented in Figure \ref{pic_mmt}. The degree of polarization is quantified by the rotated Stokes parameter $q(\lambda)=[Q_{\lambda} \cos 2\theta + U_{\lambda} \sin 2\theta]/F_{\lambda}$, where $\theta$ is the mean polarization angle of the source in the 5000-8000\AA\ bandpass, $F_{\lambda}$ is the total optical flux density, and $Q_{\lambda}$ and $U_{\lambda}$ are Stokes parameters. Unlike the conventional polarization $P(\lambda)=\sqrt{Q_{\lambda}^2+U_{\lambda}^2}/F_{\lambda}$, the value $q(\lambda)$ is not positive definite and has a normal error distribution. The polarization position angle is defined as $\tan 2\theta(\lambda)=U_{\lambda}/Q_{\lambda}$ and is measured in degrees East of North. In Figure \ref{pic_mmt}, the degree of polarization, the polarization angle and the polarized flux $q(\lambda)\times F_{\lambda}$ have been heavily binned in wavelength because of the low signal-to-noise ratio at the original spectral resolution. In Table \ref{tab:pol}, we summarize spectropolarimetric data in the form of broad-band measurements (`UV', `blue' and `yellow') obtained by error-weighting spectral data within the corresponding ACS filters' bandpasses. 

Broad-band polarimetry was obtained for 11 objects from Paper I in September 2004 (see Appendix \ref{sec_app} for the complete dataset). Among them were two objects from our HST program (SDSS~J0123+0044 and SDSS~J2358$-$0009). Observations were conducted using SPOL in its imaging mode at the Bok 2.3-m reflector on Kitt Peak, AZ. The instrumental configuration and data processing were similar to those discussed by \citet{smit02}. In brief, a semiachromatic wave plane is placed in the beam and rotated through 16 positions to determine the linear Stokes parameters, $Q$ and $U$, using exposure times for each position of the plate of 200-240 sec, with a total integration time per object of 3200-3840 sec. All observations were through a $V$ filter (4800$-$6000\AA) to prevent [OIII]5007 from being included in the bandpass. The resulting broad-band polarization values (again quoted as rotated Stokes parameters), position angles and their uncertainties are listed in Table \ref{tab:pol}. In both objects polarization is detected at less than 2$\sigma$ confidence level. 

\section{Image analysis and morphology}
\label{sec_image_analysis}

Our morphological analysis of the HST data is aimed at determining (i) whether these galaxies can be described as disks, disks with bulges, or pure ellipticals, (ii) whether there is any evidence for a non-stellar contribution to the HST images, (iii) whether there are significant morphological distortions of the hosts indicative of recent or ongoing interactions, and (iv) the luminosities and colors of the stellar populations of the host galaxies. We start by performing aperture photometry of our objects in Section \ref{sec_aper} and then proceed to derive model photometry. We summarize the definition of the S\'ersic profile in Section \ref{sec_sersic}. In Section \ref{sec_1d} we describe one-dimensional (1-D) surface brightness profiles of the galaxies in our sample. We then use the results of the 1-D analysis in fitting smooth two-dimensional (2-D) elliptical models in Section \ref{sec_2d}. 

\subsection{Aperture photometry}
\label{sec_aper}

For a given object, we used elliptical apertures with ellipticities, position angles and semi-major axes determined by the outer isophotes in the yellow image. The same aperture was used in all three bands for this object. We tried several aperture sizes and chose the ones which were large enough so that further increase in aperture size did not increase the total enclosed flux by more than 5\%, while at the same time avoiding close companions. The aperture sizes were in most cases somewhat larger than the outermost detected isophote (our typical surface brightness sensitivity is 25$-$26 mag/arcsec$^2$ in the yellow band). We estimate that the resulting aperture magnitudes (Table \ref{tab:aper}) are accurate to better than 0.1 mag. 

\subsection{S\'ersic brightness profile}
\label{sec_sersic}

If the isophotes of a galaxy are elliptical with the same ellipticity $\varepsilon=1-$(semi-minor axis/semi-major axis), the same position angle on the sky and the same center, its photometry can be completely described by the 1-D surface brightness profile $I(r)$, where $r$ is the semi-major axis of the isophote. The \citet{sers68} profile is a three-parameter function:
\begin{equation}
\log I(r)=a_0 - a_1 r^{1/n_s},
\label{eq_sersic}
\end{equation}
where $n_s$ is referred to as the S\'ersic index. The surface brightness profiles of many galaxies can be described by the S\'ersic function or a superposition of two functions with different $n_s$ \citep{binn98}. Elliptical galaxies and bulges of spirals typically have $n_s\simeq 4.0$ \citep{deva48}, while disk components typically have an exponential profile, i.e., $n_s\simeq 1.0$ \citep{patt40, free70}. In equation (\ref{eq_sersic}), parameter $a_1$ determines the scale length of the galaxy. For the exponential and the de Vaucouleurs profiles this parameter can be related to the conventionally used exponential scale length and half-light radius, correspondingly: $R_d=0.434/a_1$ and $R_e=(3.33/a_1)^4$. 

Assuming that the profile continues to infinity, the total flux emitted by a galaxy with a 1-D surface brightness profile given by equation (\ref{eq_sersic}) and with ellipticity $\varepsilon$ is
\begin{equation}
F=(1-\varepsilon)2\pi 10^{a_0} \frac{n_s \Gamma(2n_s)}{(a_1 \ln 10)^{2 n_s}}. 
\label{eq_sersic_flux}
\end{equation}
If $n_s$ is an integer (as it is for the de Vaucouleurs and the exponential profiles), then $\Gamma(2n_s)=(2n_s-1)!$ 

\subsection{1-D brightness profiles}
\label{sec_1d}

\begin{figure}
\epsscale{0.8}
\plotone{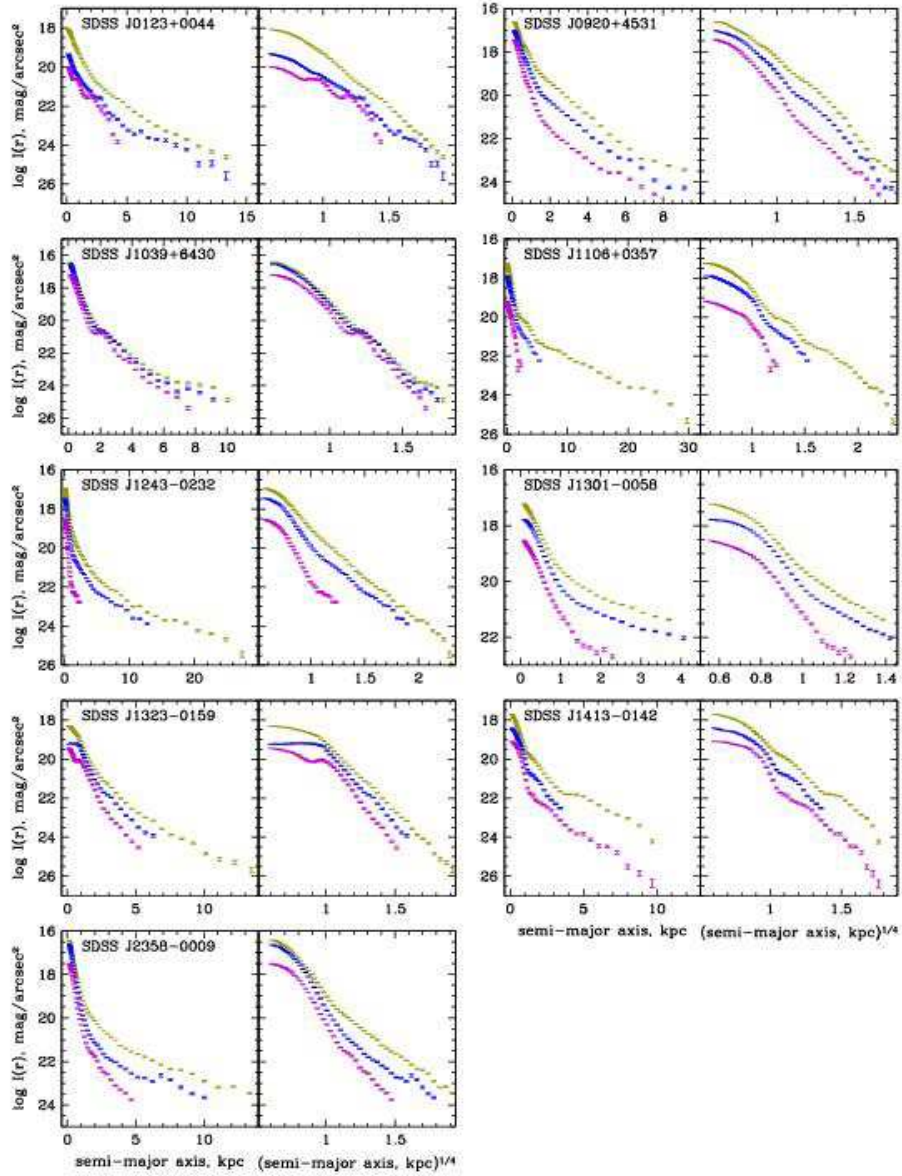}
\caption{1-D brightness profiles of the extended emission, regardless of its origin, for all objects in all bands. For each object, $\log I(r)$ is plotted in units of apparent AB mag/arcsec$^2$ as a function of the semi-major axes of the isophotes $r$ and as a function of $r^{1/4}$. Yellow band brightness profiles are shown in yellow, blue band in blue and UV in purple. The error bars reflect the rms scatter of the values of the surface brightness on the fitted isophote. The routine fails in the outer parts of SDSS~J1301$-$0058 because of the presence of a nearby companion, so the profiles are shown only for the inner 4 kpc. The UV filters used for SDSS~J1106+0357, SDSS~J1243$-$0232 and SDSS~J1301$-$0058 were significantly less sensitive than those used for other objects, so the UV 1-D profiles of these objects cover only the well-detected inner parts.}
\label{pic_iso_params}
\end{figure}

\begin{figure}
\epsscale{0.9}
\plotone{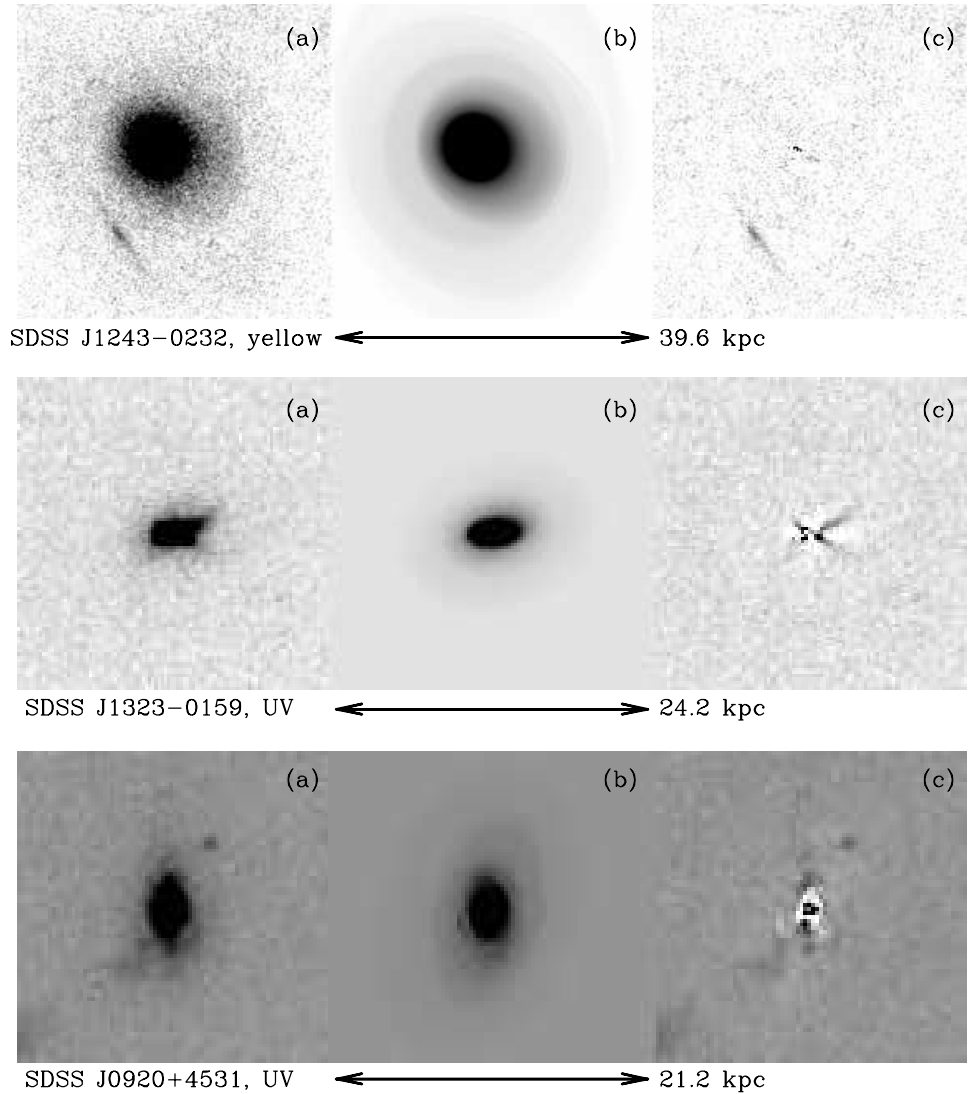}
\caption{Examples of fitting elliptical isophotes for three objects. From left to right: (a) the image, (b) the model reconstructed from the best-fit elliptical isophotes, (c) residuals. As seen in the top panel, residuals in the yellow are 1\% or less except for a few central pixels, and are typically much lower than those in the UV (bottom two objects) where they reach tens of per cent. All black-and-white images in this paper are negatives: darker areas mean more flux per pixel. Orientation of these and subsequent images is the same as in Figures \ref{pic_rgb1}-\ref{pic_rgb2}.}
\label{pic_isophote}
\end{figure}

For each object and each band, we use the standard IRAF package {\sl ellipse} to fit isophotes with ellipses, without regard to whether the light is due to the stellar or the scattered component. Isophotes are fitted at pre-defined semi-major axes using the iterative algorithm described by \citet{jedr87}. For each semi-major axis, the output includes the best-fit surface brightness, center, ellipticity, position angle and other parameters. We then plot the surface brightness as a function of the semi-major axis, without taking into account changes in the centers, ellipticities or position angles from one isophote to another. The resulting 1-D profiles are shown for all objects in all filters in Figure \ref{pic_iso_params} both as a function of $r$ and $r^{1/4}$. In most cases (especially in the yellow and blue bands) the models constructed based on best-fit elliptical isophotes reproduce the observed brightness profile very well (better than 1\%), but in some cases (typically in the UV) significant residuals are present; examples are shown in Figure \ref{pic_isophote}. 

The point-spread function (PSF) suppresses the central flux and redistributes it outward, causing the flattening of the inner parts of each profile (especially apparent in the $r^{1/4}-\log I(r)$ plots in Figure \ref{pic_iso_params}). The PSF also circularizes the isophotes and slightly increases the effective radii. We explicitly incorporate the PSF in our 2-D fits (Section \ref{sec_2d}). 

Although the constructed 1-D brightness profiles do not take into account the changes in the ellipticities, position angles or positions of the centers of the isophotes with semi-major axis (and neither do they discriminate between scattered and stellar components), they provide some insights into the structures of the galaxies and a starting point for further analysis (Section \ref{sec_2d}). Figure \ref{pic_iso_params} shows that the de Vaucouleurs profile fits many of the galaxies very well. Strong evidence for an exponential component is present only in SDSS~J1106+0357 (which indeed looks like a spiral galaxy, Figure \ref{pic_rgb2}). The 1-D profiles of SDSS~J0920+4531 and SDSS~J1413$-$0142 show changes in slope, possibly indicating a transition from a bulge to a disk component (although these changes do not occur consistently in all three bands). In some objects, e.g. SDSS~J1323$-$0159, there is a clear color gradient that can be seen as different slopes of the brightness profiles in three bands -- in this case, the UV light is more centrally concentrated than blue or yellow. 

A number of other features can be identified in the 1-D brightness profiles. For example, the prominence of wiggles, such as those seen in the 1-D profiles of SDSS~J0920$+$4531 and SDSS~J1413$-$0142, increases toward the blue, suggesting that they are due to scattered light or patches of star formation. Some centers are clearly underluminous compared to the best-fit smooth profiles (e.g., SDSS~J1323$-$0159), while others show a strong central excess of light despite the smoothing due to the PSF (e.g., SDSS~J1301$-$0058). We will argue that flux deficiencies (seen as negative residuals when the best-fit 2-D model is subtracted) are caused by dust obscuration, whereas the interpretation of flux excesses (seen as positive residuals) may be due to scattered light, patches of star formation or a superposed companion galaxy.   

\subsection{2-D brightness profiles}
\label{sec_2d}

In Paper II, we demonstrated that at least in some cases light emitted by the central AGN and scattered off interstellar material in the host galaxy contributes a significant portion of the total flux and therefore should be present in the HST images. In Figures \ref{pic_rgb1}-\ref{pic_rgb2} we show the orientations of the brightest UV spots or cones and the orientation of the polarization vector (where polarimetric data are available). In some cases (such as SDSS~J1323$-$0159 and SDSS~J1413$-$0142 described in Paper II) the scattering regions can be identified based on their morphological similarities to well-studied nearby examples \citep{pogg93}, and this identification agrees with the orientation predicted by spectropolarimetric measurements. In most cases, the polarization position angle is orthogonal to the line between the centers of the galaxies and the presumed scattering regions, but there are exceptions (such as SDSS~J0920+4531, where the brightest UV spot does not seem to be responsible for the polarization) and complications (such as SDSS~J0123+0044, where numerous spots are detected). In addition, the polarization signal is suppressed when scattering occurs along multiple directions, partially canceling the polarization vectors; in this case the scattering regions cannot be unambiguously identified from the polarimetric data. 

The major complication of our analysis is distinguishing between scattered and stellar components. In the absence of detailed imaging polarimetry some assumptions must be made about the two components to allow their separation. We fit our galaxies with a 2-D de Vaucouleurs model (elliptical galaxy), or a sum of a de Vaucouleurs model and an exponential model (disk galaxy with a bulge) in the cases where the presence of a disk is suggested by the form of the 1-D profile (Section \ref{sec_1d}). We assume that the stellar component of the host galaxy is well represented by these simple models, whereas the residuals are comprised of scattered light, isolated regions of star formation and dust obscuration. We comment on the robustness of the scattered light identification for each object in Section \ref{sec_ind}. For each object, the images in the three bands are modeled independently. With the exception of UV-band images of group 2 objects, the galaxies are typically detected well outside two effective radii, and out to five or so effective radii in some cases, allowing robust fitting of the outer parts. Our adopted models are certainly overly simplified, but the quality of our data and the presence of strong irregular morphological features prevents the use of more complex models with a larger number of free parameters (for an extensive library of galaxy models see, e.g., \citealt{peng02}).

We convolve all models with the PSF before fitting the data. We used {\sl TinyTim}\footnote{http://www.stsci.edu/software/tinytim} to generate PSFs for all filters in our observations. For the broader filters, the PSF has some dependence on the assumed spectral energy distribution within the filter. In these cases, we used an approximation to the spectrum in the form $F_{\lambda}=$const. The size of the PSF kernel used in the convolution is 3\arcsec. 

In the simplest case of a de Vaucouleurs model, there are six fitting parameters: total normalization $a_0$, slope $a_1$, ellipticity, the position of the center (two coordinates), and the position angle of the major axis. We assume concentric, aligned isophotes with ellipticities that are constant with the distance from the center. In the bulge$+$disk model, there are 12 parameters (6 for each of the S\'ersic components) that are allowed to vary independently. The mean value of the sky is kept constant during 2-D fitting after it is determined by fitting a Gaussian to the distribution of pixel values for a patch of blank sky in the same image. The value being minimized to find the best fit is $\sum_i(F_i-M_i)^2/\sigma_i^2$, where summation is over all unmasked pixels, $F_i$ is the flux of the $i$-th pixel, $M_i$ is the model value of the same pixel and $\sigma_i$ is an estimate of the flux error per pixel that combines the Poisson noise and the read-out noise. 

The most challenging part in fitting the 2-D profiles to the data is to identify and mask out regions with significant deviations from the S\'ersic profiles. Polarization data help identify the major scattering regions as demonstrated in Figures \ref{pic_rgb1}-\ref{pic_rgb2}. We used an iterative procedure in which the fitting residuals are used to construct a mask for the next round of fitting, and the procedure is repeated until the residuals very closely resemble the mask. In some cases the results of the fitting are sensitive to the assumed mask, leading to large uncertainties in structural parameters such as half-light radii or ellipticities. However, the errors in these parameters are highly covariant, with the result that the uncertainty in the measurement of the integrated flux is rather small. We estimate errors by running a series of 2-D fits with different masks, selecting acceptable fits and looking at the range of structural parameters and luminosities that they yield. 

In Table \ref{tab:par} we list the best-fit structural parameters that resulted from our 2-D modeling. The absolute magnitudes listed in Table \ref{tab:par} are obtained using equation \ref{eq_sersic_flux} (`model magnitudes' hereafter), i.e., assuming that the best-fit profiles extend to infinity. 

Whenever our fitting procedure results in negative residuals, these residuals typically increase toward shorter wavelengths, so it is natural to attribute them to dust obscuration. In these cases, we calculate the ratio of observed to model flux in the yellow band (which is close to the rest-frame $V$ band in all cases) and correct it to the $V$ band using an extinction law in the form $A_{\lambda} \propto 1/\lambda$ (e.g., \citealt{card89}), so that all values of extinction are given corrected to the rest-frame $V$-band. These corrections are much smaller than the uncertainties in the model.

In order to compute the ratio of the scattered to stellar light listed in Table \ref{tab:par}, we first calculate the total stellar luminosity using equation \ref{eq_sersic_flux} and then perform aperture photometry on the positive residuals that we confidently attribute to scattered light based on the polarization measurements. In Section \ref{sec_ind}, we explicitly describe which regions were included for each object. Our procedure is likely to underestimate the fractional importance of the scattered light because we include only those positive residuals that are unambiguously due to scattering, and because low surface brightness scattered emission may disappear in the fit, may be underestimated because of dust obscuration or may simply remain undetected. 

\section{Individual objects}
\label{sec_ind}

\subsection{SDSS~J0123+0044, $z=0.399$}

\begin{figure}
\epsscale{0.9}
\plotone{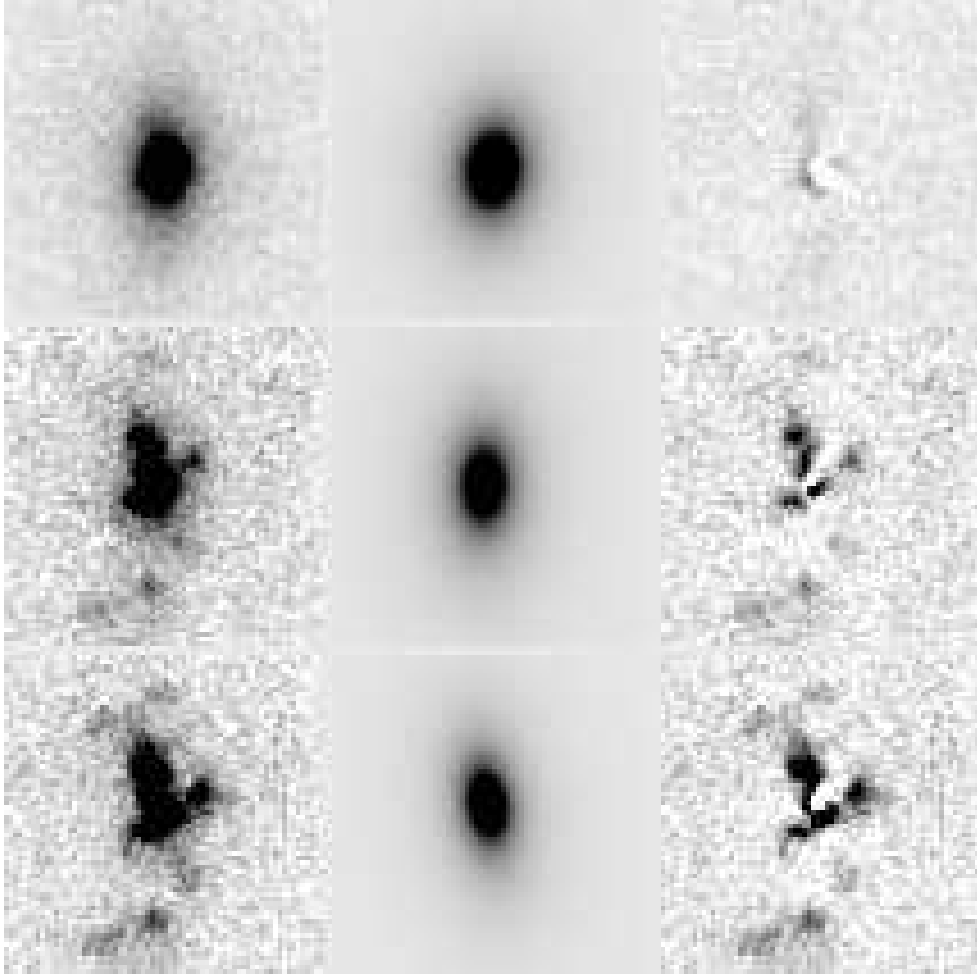}
\caption{Subtracting a 2-D de Vaucouleurs profile from the image of SDSS~J0123+0044. From top to bottom: yellow, blue and UV bands, on the same scale (the size of each frame is 70 pixels, or 18.4 kpc). From left to right: the image, the best-fitting model, and the residuals. Residuals become stronger toward shorter wavelengths, and the uncertainty of the model increases. In particular, the differences in the position angles and ellipticities between the bands are within the model uncertainties. Both positive and negative residuals are present; the latter appear as solid white. While negative residuals are probably due to dust obscuration, the nature of the positive residuals is less certain. }
\label{pic_2d_0123}
\end{figure}

\begin{figure}
\epsscale{0.8}
\plotone{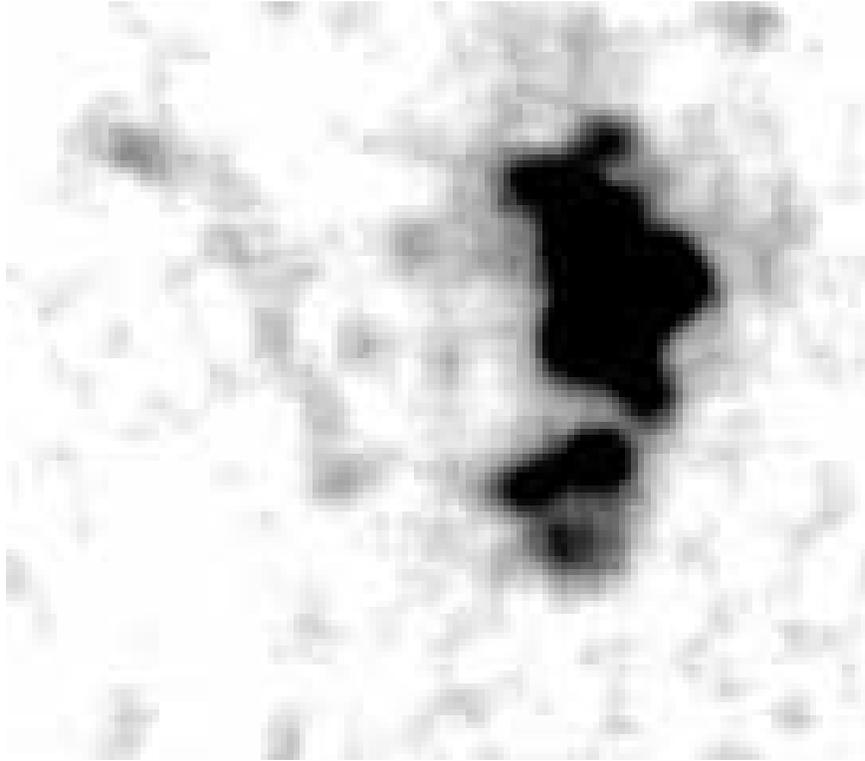}
\caption{Faint extended emission associated with SDSS~J0123+0044 in the UV band (the tail on the left-hand side of the image; the feature is not detected in the yellow and blue bands). The image has been smoothed with an 8$\times$8 boxcar filter for display purposes. The horizontal scale of the frame is 35.8 kpc. }
\label{pic_tidal3}
\end{figure}

This object is represented by a single de Vaucouleurs profile quite well, as shown in Figure \ref{pic_2d_0123}. Residuals become stronger toward shorter wavelengths, and some dust obscuration ($A_V \la 0.2$ mag) is present near the center. The nature of the residuals is ambiguous. Some of the bright UV spots in the composite image (Figure \ref{pic_rgb1}; also seen as residuals in Figure \ref{pic_2d_0123}) may be scattering regions, since the line connecting the brightest spot with the nominal center of the galaxy is orthogonal to the polarization position angle. This interpretation should be regarded as somewhat uncertain, since the polarization was measured at less than $2\sigma$ confidence level. On a larger scale, the galaxy shows a faint tidal tail-like structure (Figure \ref{pic_tidal3}) indicating possible interactions. Therefore, it is likely that at least some of the residuals are star forming regions, especially those to the east of the nucleus that connect to the tidal tail. In our estimate of the scattered light contribution given in Table \ref{tab:par}, we do not include any of the positive residuals to the east of the nucleus.

\subsection{SDSS~J0920+4531, $z=0.402$}

This object shows some evidence for an exponential component in its 1-D profile, so we fit it with a bulge+disk combination, producing rather poor fits spanning a large range of possible fitting parameters depending on the chosen masks. The presence of dust obscuration and strong positive residuals with complicated morphology make our best fits particularly uncertain in this object. Dust obscuration is estimated at $A_V\la 0.5$ mag; it can be seen as a red spiral-like structure in the color-composite image in Figure \ref{pic_rgb1} to the east of the nucleus, suggesting that the eastern side is closer to the observer. We estimate that the uncertainty in the measurement of the total stellar magnitude in this object is about 0.3 mag, compared to $<$0.1 mag in the objects with better fits. This is the only object in the sample in which the brightest UV region in the color-composite image (the one to the north-west of the nucleus schematically shown as a grey cone in Figure \ref{pic_rgb1}) is not orthogonal to the measured polarization position angle. In fact, the presence of at least three morphologically irregular companions within 25 projected kpc and the morphology of the UV region suggests that it might be a superposed companion galaxy with on-going star formation. We therefore do not make an estimate of the scattered-to-stellar ratio for SDSS~J0920+4531, since the scattering regions cannot be confidently identified. 

\subsection{SDSS~J1039+6430, $z=0.402$}
\label{sec_1039}

\begin{figure}
\epsscale{1.0}
\plotone{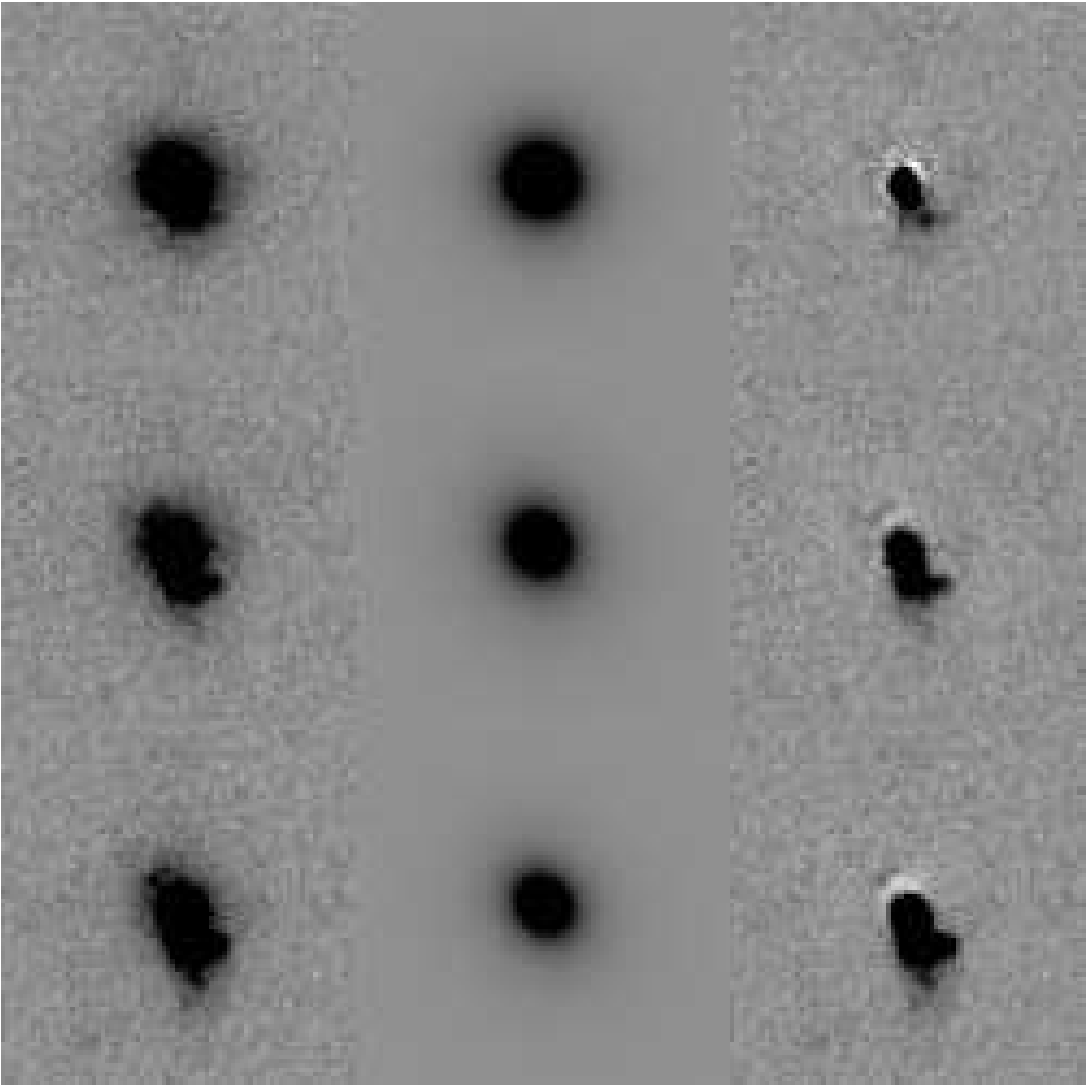}
\caption{Subtracting a de Vaucouleurs profile from the image of SDSS~J1039+6430. The frames, which are 82 pixels (21.6 kpc) wide, are in the same order as in Figure \ref{pic_2d_0123}. Almost all residuals are positive and are due to scattered light. }
\label{pic_2d_1039}
\end{figure}

We model the stellar light in this object with a single de Vaucouleurs profile, as shown in Figure \ref{pic_2d_1039}. We interpret all of the positive residuals as scattered light, taking into account the very high measured polarization in this source, which reaches 17\% in the UV (Paper II). Some dust obscuration ($A_V\la 0.3$ mag) may be present to the north of the nucleus. After computing the scattered-to-stellar ratio using aperture photometry of the positive residuals, we correct the observed values of polarization in all three bands (Table \ref{tab:pol}) for the contribution of the stellar light (assumed unpolarized). We find the polarization of the scattered component to be 26$\pm$3\%, independent of wavelength within the errors. An additional correction can be made to account for the finite width of the scattering cone: particles at different positions within the cone produce polarization at somewhat different polarization position angles, leading to partial cancellation of polarization. If the opening angle of a homogeneously-filled scattering cone is $\theta_0$, the effect reduces the intrinsic polarization by a factor of $\sin\theta_0/\theta_0$. For SDSS~J1039+5430, $\theta_0\simeq 50$\degree, so our final estimate for the intrinsic polarization is 30\%. 

The spectacular large-scale (10 kpc in projection) scattering region to the south of the center was discussed in Paper II. Based on the large size of the scatterer, we argued that dust rather than electron scattering produces the high observed polarization in this object. This argument is further supported by the one-sidedness of the scattering region; although its counterpart to the north of the nucleus is detected and can be seen as a bluish spot on the color-composite image (Figure \ref{pic_rgb1}) and in the residuals (Figure \ref{pic_2d_1039}), it is much fainter. The difference in the brightness of the northern and the southern spots can be explained due to the preferentially forward scattering of dust particles, if the axis of the scattering cones points somewhat toward the observer in the southern part. 

\subsection{SDSS~J1106$+$0357, $z=0.242$}

\begin{figure}
\epsscale{1.0}
\plotone{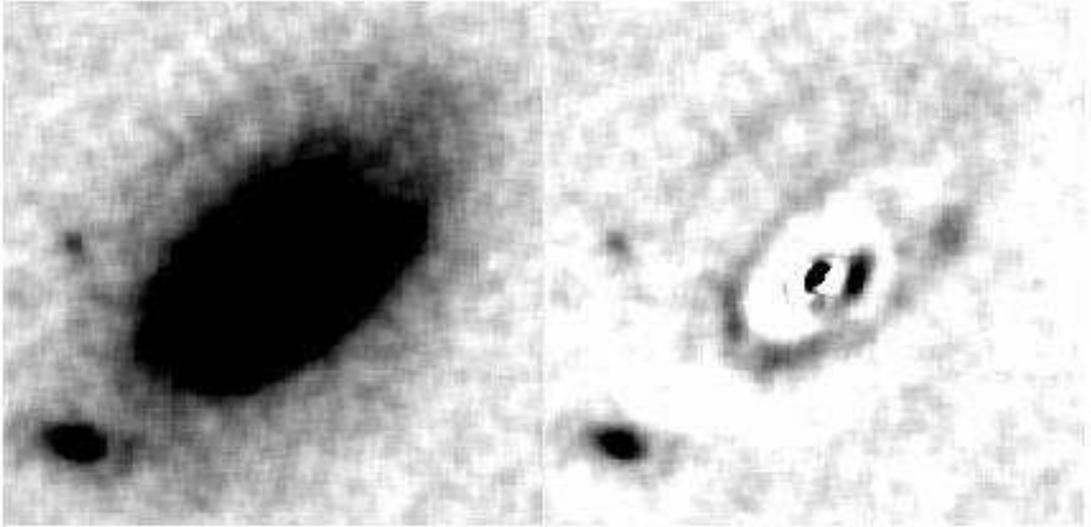}
\caption{Subtracting a sum of a de Vaucouleurs profile and exponential profile from the image of SDSS~J1106$+$0357 in the yellow band. The frames, which are 254 pixels (48.5 kpc) wide, show the original image and the residuals after subtraction. Both frames have been smoothed with an 8$\times$8 boxcar to highlight the low surface brightness features. A faint extended arm connecting the main galaxy to an extended companion to the north-east (image left) of the nucleus can be seen, especially in the residual image.}
\label{pic_2d_1106}
\end{figure}

This object is the largest and best-measured bulge+disk galaxy in our sample. In the yellow and blue bands, significant deviations from the best fitting bulge+disk are present (Figure \ref{pic_2d_1106}). Obscuration is present both in the outer disk ($A_V\la 1$ mag) and in the central parts of the galaxy ($A_V\la 0.6$ mag). In the outer disk the obscuration is preferentially confined to the northern (image top left) side, suggesting that this side is closer to the observer. The scattering region is well defined and has a classical biconical shape, and it is accurately orthogonal to the measured polarization position angle. Although the scattering regions are detected out to 1.5 kpc from the center, they are dwarfed by the galaxy which has a disk exponential scale well above 10 kpc (Table \ref{tab:par}). The disk component is not detected in the UV-band image, presumably because of the low sensitivity of the ramp filter used in this observation, and in Table \ref{tab:par} we give a lower limit on the UV-band stellar luminosity of the bulge component only. In the yellow image, at least three lumps can be seen outside the main disk but within 25 kpc (projected) of the center of this galaxy. A faint arm (seen in Figure \ref{pic_2d_1106}) connects the outskirts of the disk with an extended companion to the north-east (image left) of the center. If this feature lies in the plane of the galaxy and if the observed ellipticity of about 0.4 is entirely due to inclination, then this arm follows a nearly circular arc around the center of this galaxy, with a radius of 25 kpc. 

\subsection{SDSS~J1243$-$0232, $z=0.281$}

\begin{figure}
\epsscale{1.0}
\plotone{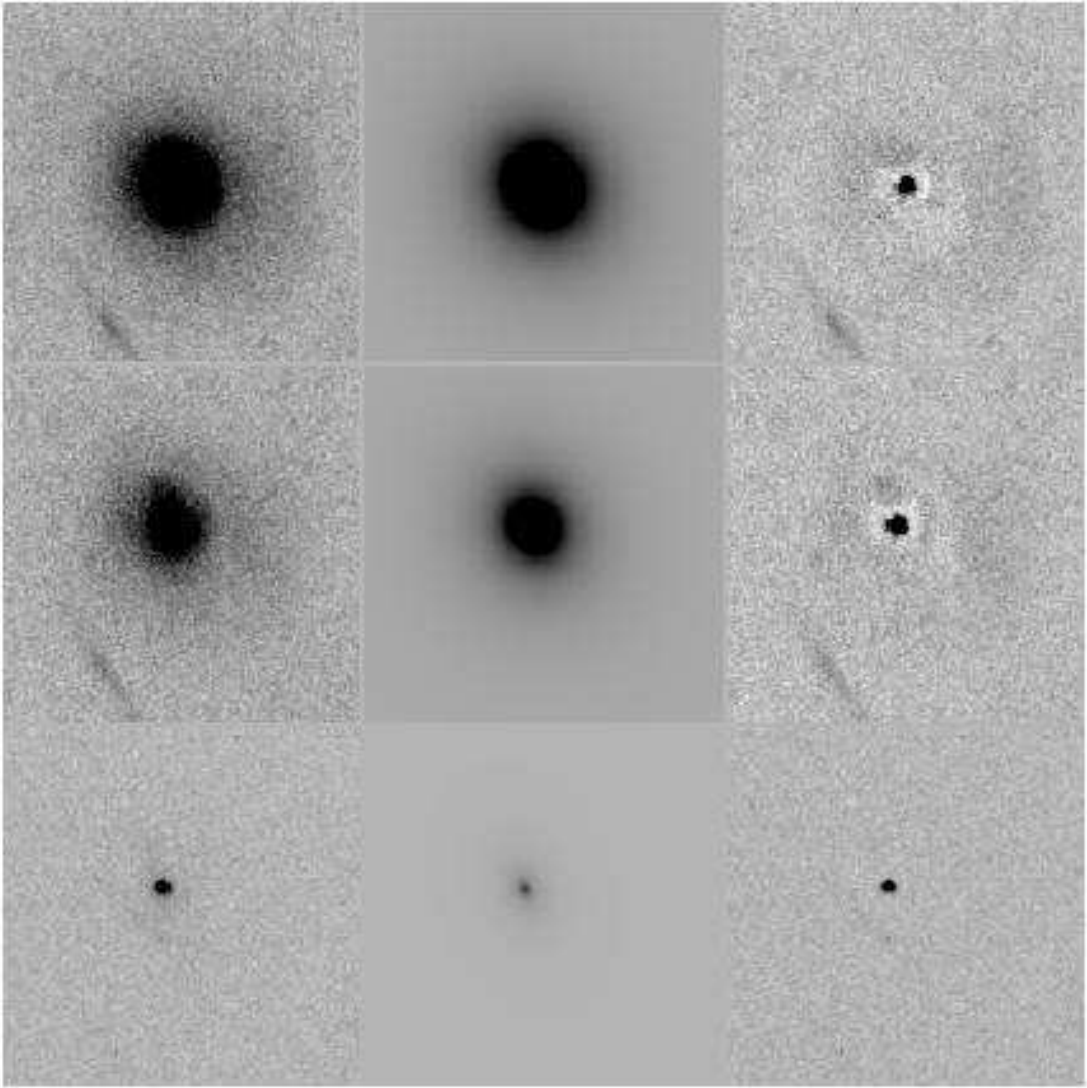}
\caption{Subtracting a de Vaucouleurs profile from the image of SDSS~J1243$-$0232. The frames, which are 130 pixels (27.1 kpc) wide, are in the same order as in Figure \ref{pic_2d_0123}. There is excess emission in the center relative to the best-fit de Vaucouleurs profile; this excess is also apparent in the 1-D profile of this object (Figure \ref{pic_iso_params}). This object is one of the sources which was imaged using the low-sensitivity ramp filter FR459M (Table \ref{tab:ids}, group 2), and the extended UV emission is just barely detected. }
\label{pic_2d_1243}
\end{figure}

This object is very well represented by a de Vaucouleurs profile. However, there is a significant central flux excess over the best-fit de Vaucouleurs component within the few central pixels (Figures \ref{pic_iso_params} and \ref{pic_2d_1243}). The excess is very blue in color, approximately $F_{\lambda}\propto \lambda^{-1.5}$, but whether this excess is due to a circumnuclear starburst or to sub-kpc scattering is not clear. This excess is excluded from the calculation of the model magnitudes.

\subsection{SDSS~J1301$-$0058, $z=0.246$}

We start by analyzing the largest component in this guitar-shaped interacting galaxy. As in SDSS~J1243$-$0232, the 1-D profile of this galaxy shows a clear central excess over the de Vaucouleurs fit to the outer parts of this object. A combination of two de Vaucouleurs profiles (one representing the central excess and one representing the outer parts) provides the best fit to the 2-D images. The more extended de Vaucouleurs component is clearly of stellar origin, but the nature of the more compact component ($R_e\simeq 0.6$ kpc) is unclear. As for SDSS~J1243$-$0232, we only include the data for the more extended of the two components in the calculation of the luminosity of the stellar component (listed in Table \ref{tab:par}). Some residuals are present when the best fit is subtracted. The orientation of these residuals is schematically shown in Figure \ref{pic_rgb2}, but since no polarization data are available for this object, we cannot determine whether if these residuals are caused by scattered light. The companion galaxy to the south-west (lower left) of the main object is well-represented by a disk+bulge combination and shows UV bow-like residuals in the center that could also be interpreted as scattering cones. 

It is not clear which of the two objects hosts the type II quasar. The separation between the galaxies is 1.3\arcsec, so that they both comfortably fit into the 3\arcsec\ diameter SDSS fiber aperture. Both galaxies have irregular central parts, so given the range of morphologies of the scattered light in our sample, either of the two objects (or both) could be the actual AGN host.

\subsection{SDSS~J1323$-$0159, $z=0.350$}

\begin{figure}
\epsscale{1.0}
\plotone{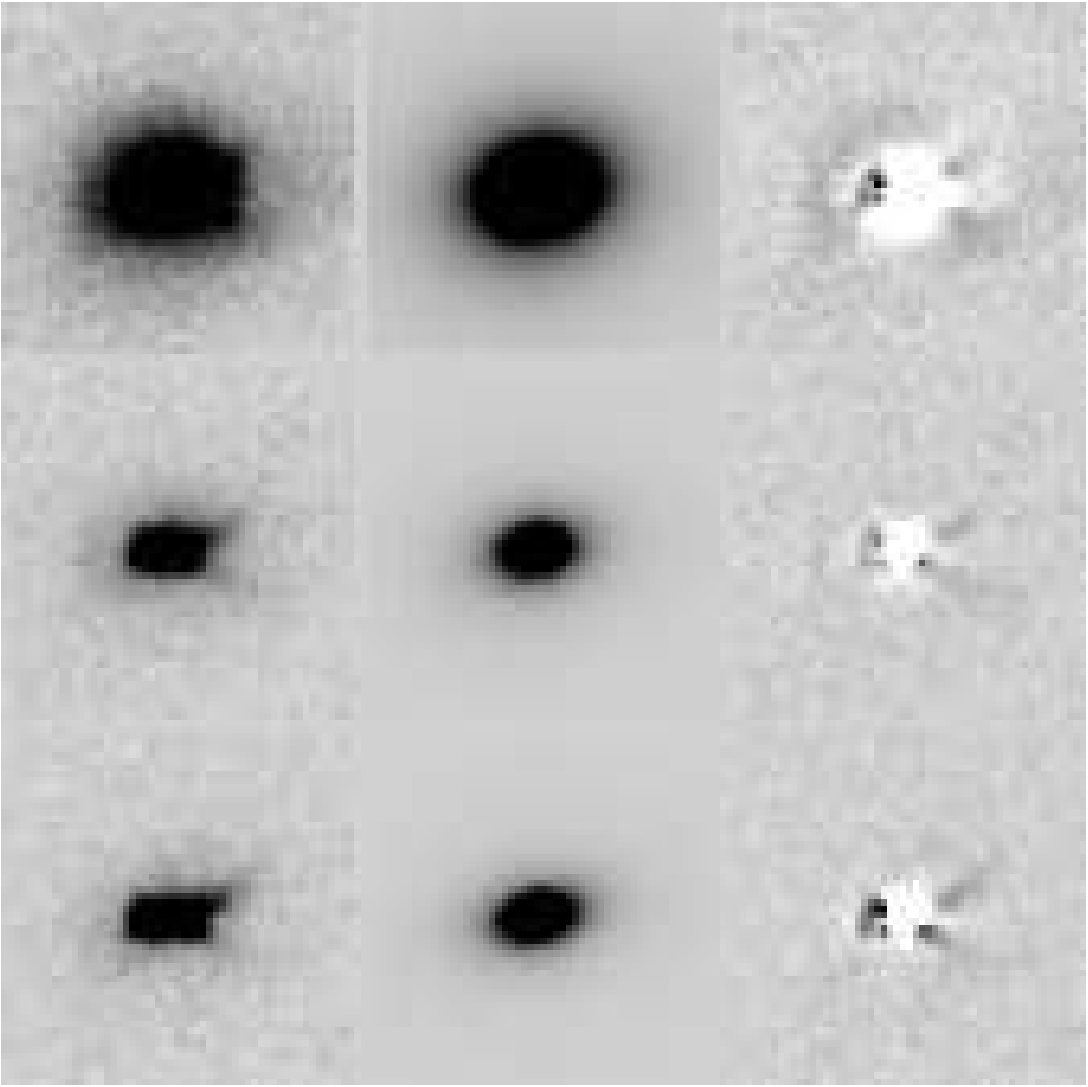}
\caption{Subtracting a de Vaucouleurs profile from the image of SDSS~J1323$-$0159. The frames, which are 55 pixels (13.3 kpc) wide, are in the same order as in Figure \ref{pic_2d_0123}. The scattering cones (X-shaped structures) are apparent in the residuals. The central flux is strongly suppressed relative to the best-fitting de Vaucouleurs profile, in agreement with the results of the 1-D analysis (Figure \ref{pic_iso_params}). This is indicative of a kpc scale obscuration in this object, which is also suggested by the color-composite image (Figure \ref{pic_rgb1}). }
\label{pic_2d_1323}
\end{figure}

The outer parts of this galaxy are well-fit by a de Vaucouleurs profile (Figure \ref{pic_2d_1323}), but there are significant non-axisymmetric negative and positive residuals in the center, making 2-D fitting of this object quite challenging. These difficulties are manifested by the range of the effective radii allowed by our fits shown in Table \ref{tab:par}. Nevertheless, the uncertainties in the measured absolute magnitudes are quite small ($<$0.1 mag) in the yellow and blue bands. In the UV, the uncertainties increase to 0.2 mag because the stellar light is more compact in this band and the fit depends more sensitively on the assumed central mask. 

This object was first presented in Paper II, where we argued, on the basis of morphology, that the UV X-shaped structure (seen as positive residuals in Figure \ref{pic_2d_1323}) is a biconical scattering region. This interpretation is supported by the fact that the axis of this structure is orthogonal to the polarization position angle. The negative residuals can be seen as a red streak going vertically through the center of the color-composite image of this object (Figure \ref{pic_rgb1}) and can be interpreted as a kpc scale dust lane with the largest observed extinction about $A_V\simeq 1.2$ mag.

\subsection{SDSS~J1413$-$0142, $z=0.380$}

The 1-D surface brightness profiles of this galaxy show evidence for two breaks (at 1.5 kpc and at 4.5 kpc in semi-major axis), and the profile between the breaks is roughly exponential. In our 2-D fitting procedure, we fit this object with a disk+bulge combination. As for SDSS~J0920$+$4531, this results in rather poor fits. 

This object was first presented in Paper II and shows two bright conical structures (schematically represented by grey cones in Figure \ref{pic_rgb1}) roughly orthogonal to the polarization position angle, although not quite as symmetric around the center as the more classical cones of SDSS~J1323$-$0159. In addition to these prominent structures, excess light is seen in the residuals and in the color composite to the south and to the north of the nucleus. Furthermore, residuals show that some extinction is present in the central region. As a result, the fits for the bulge are highly uncertain, especially in the UV band. In our two-component model of the galaxy, the disk is offset relative to the bulge by about 0.4 -- 0.65 kpc from the center along the semi-major axis, roughly toward the south of the nucleus (this asymmetry can be seen in Figure \ref{pic_rgb1}). In the estimate of the scattered flux, we include all of the positive residuals because of their conical morphology. 

\subsection{SDSS~J2358$-$0009, $z=0.402$}

The 1-D profile of this object suggests a de Vaucouleurs profile, which produces a good 2-D fit, but with strong positive central residuals with close-to-zero ellipticity. 2-D modeling of the UV image produced a significantly larger effective radius than that suggested by the 1-D profile, and therefore the UV effective radius is not listed in Table \ref{tab:par}. The ambiguity is due to the fact that most of the detected UV light is masked out from the fit, as it is part of the central excess rather than the extended stellar light. It is not clear whether these residuals are due to scattered light or a circumnuclear starburst. In addition, a narrow UV cone extending to the north-east of the nucleus and orthogonal to the polarization position angle (Figure \ref{pic_rgb1}) is well-detected in the residuals, and we interpret this feature as scattered light and include it in Table \ref{tab:par}, although we caution that the polarization is detected at less than 2$\sigma$ confidence level in this object. Curiously, the UV cone is aligned with a tidal structure connecting SDSS~J2358$-$0009 to the companion galaxy, so another possible interpretation is a stretched star-forming region. 

Because the UV feature appears jet-like, we also test the possibility that it represents synchrotron radiation from an energetic outflow, using 1.4 GHz radio data from the FIRST survey \citep{beck95, whit97} and the NVSS \citep{cond98}. There is neither a FIRST nor an NVSS detection at the position of SDSS~J2358$-$0009, and using the rms noise in the FIRST image of the field we place a 5$\sigma$ upper limit of 0.7 mJy on the flux from this source (assuming a point source). From our HST data, a lower limit on the UV flux density from the conical feature (excluding the central part where the morphology of the positive residuals is uncertain) is $F_{\lambda}\sim 4.2\times 10^{-19}$ erg sec$^{-1}$ cm$^{-2}$ \AA$^{-1}$. Therefore, the radio-to-optical index of this feature $\alpha$ (defined as $F_{opt}/F_{radio}=(\nu_{opt}/\nu_{radio})^{\alpha}$) is constrained to be $>-0.6$ by our data. While values around $\alpha=-0.5$ are not uncommon for the total emission (including jets, lobes, radio cores, host galaxy) from radio-loud sources, the jets have spectra with typical values of $\alpha$ around $-1$ \citep{meis96, samb04, jest05}. Therefore, the observational limit on $\alpha$ makes SDSS~J2308$-$0009 an unlikely jet candidate. 

\begin{figure}
\epsscale{0.8}
\plotone{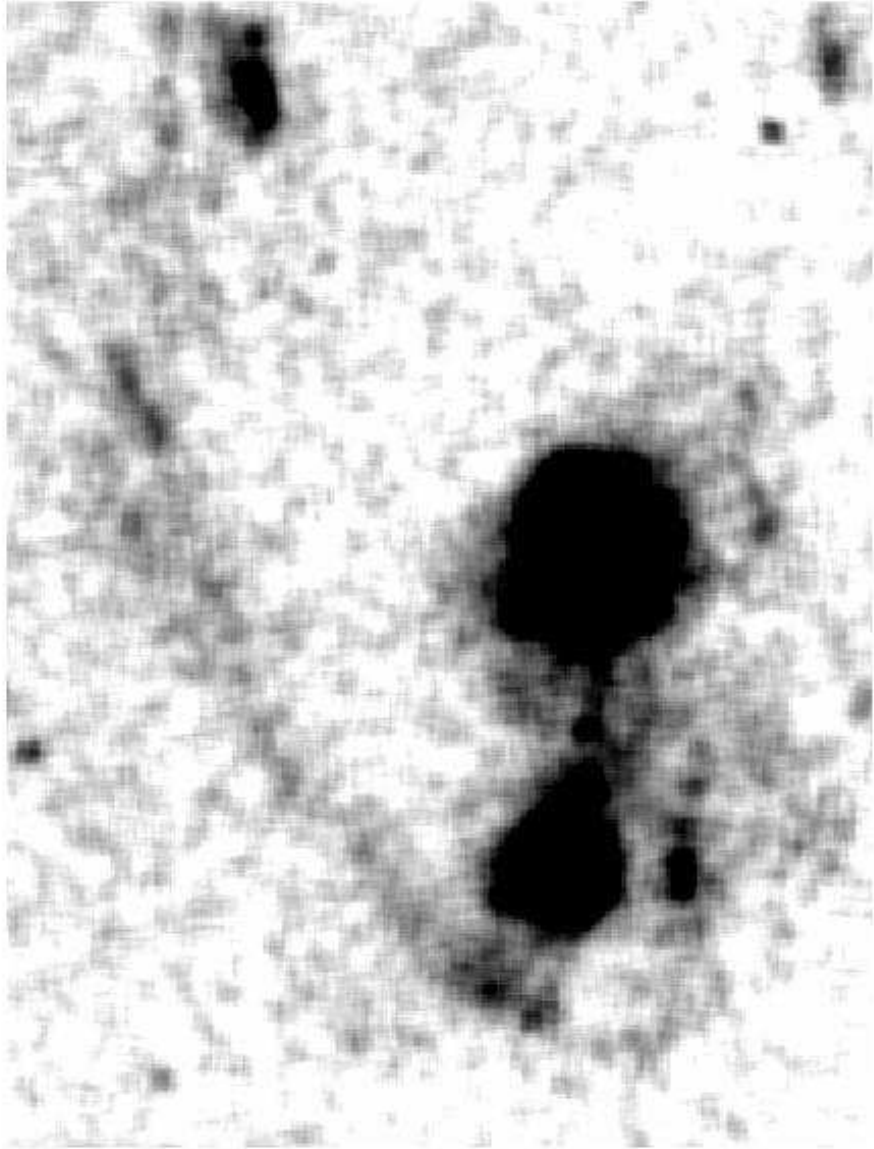}
\caption{Faint extended emission associated with SDSS~J2358$-$0009 in the yellow band. The image has been smoothed with an 8$\times$8 boxcar. The horizontal scale of the frame is 82.4 kpc. The two black blobs are the two galaxies visible in Figure \ref{pic_rgb1}.}
\label{pic_tidal5}
\end{figure}

The galaxy is clearly experiencing some type of interaction, showing large-scale tidal tails (Figure \ref{pic_tidal5}) as well as smaller scale shell-like structures to the east of the nucleus. The interacting companion to the north-east (at the projected distance of 29 kpc) is well represented by a sum of disk and de Vaucouleurs component. 

\section{Discussion}
\label{sec_discussion}

\subsection{Stellar populations of the type II quasar hosts}
\label{sec_disc1}

In this section we discuss the luminosities and colors of the stellar population of the type II quasar hosts we have imaged using the model magnitudes and characteristic radii (Table \ref{tab:par}) that resulted from our 2-D fitting procedure. To compare the luminosities of the hosts in our sample to those of the galaxies from the general population at the same redshift, we calculate the so-called $M_*$ magnitudes defined as the position of the break of the \citet{sche76} fit to the luminosity function of galaxies. The number density of galaxies with magnitudes brighter than $M_*$ decreases roughly exponentially with luminosity, so $M_*$ quantifies the luminous end of the luminosity function; \citet{bahc97} estimate that the average luminosity of a field galaxy is 1.8 mag fainter than $M_*$. We used the luminosity function of galaxies from \citet{blan03} at five optical wavelengths. We corrected their values of $M_*$ to match the redshifts of our objects using their own best-fit redshift evolution parameters. We then used linear interpolation to calculate $M_*$ values at the effective rest-frame wavelengths of our observations listed in Table \ref{tab:ids}. The resulting $M_*$ values are presented in Table \ref{tab:par} in parentheses next to the model magnitudes of our galaxies. Although the $M_*$ values by \citet{blan03} have nominal accuracies of $\le 3\%$, our calculation of $M_*$ involves extrapolation of their results to higher redshifts ($z\le 0.4$) than those of the galaxies used in constructing the luminosity functions ($z<0.22$), so we quote only three significant digits for our computed $M_*$ in Table \ref{tab:par}. 

\begin{figure}
\epsscale{0.9}
\plotone{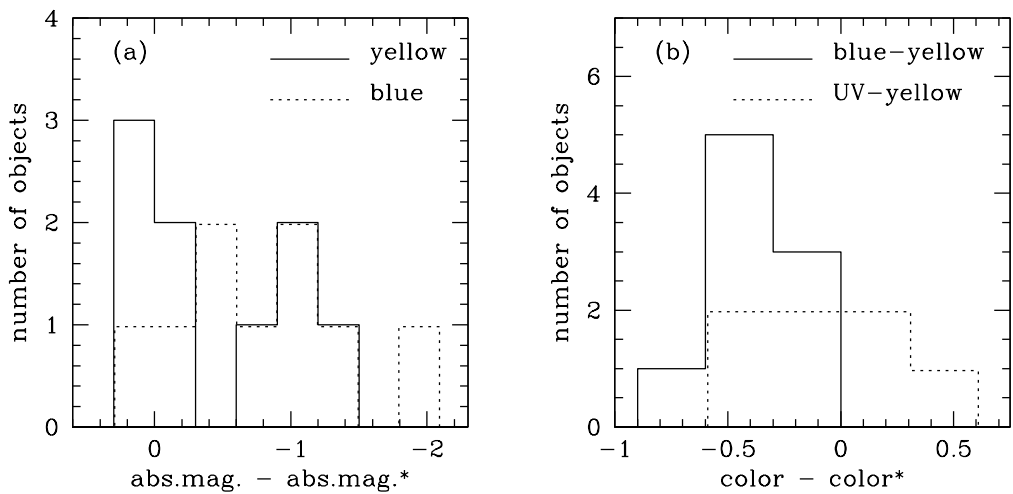}
\caption{(a) Distribution of absolute luminosities of type II quasar hosts relative to the $M_*$ values in the yellow band (solid; median value $-0.3^{+0.2}_{-0.4}$ mag; bootstrap error on the median was used) and in the blue band (dotted; median value $-0.7^{+0.2}_{-0.4}$ mag). Values $<0$ mean brighter than $M_*$. (b) Distribution of colors of type II quasar hosts relative to the $M_*$ values: blue$-$yellow colors (solid; median value $-0.4$ mag) and UV$-$yellow colors (dotted; median value $-0.2$ mag). Relative colors $<0$ mean bluer colors than those of $M_*$ galaxies. }
\label{pic_lum_col}
\end{figure}

In Figure \ref{pic_lum_col}a we compare the absolute model magnitudes of the stellar components of out type II quasar hosts with $M_*$ values. This comparison shows that in the yellow band five of the nine objects are within 0.3 mag of $M_*$, while the remaining four are 0.7$-$1.4 mag more luminous. The median difference is $(M_{host}-M_*)_{med}=-0.3$ mag in this band. The luminosity difference between type II quasar hosts and field galaxies is more pronounced in the blue band, in which type II quasar hosts are up to 2 mag more luminous than the $M_*$ values, with a median difference of 0.7 mag. Similarly, the hosts of unobscured quasars are often found to be significantly more luminous than are field galaxies, but quantitative comparison is complicated. For example, \citet{bahc97} find $(M_{host}-M_*)_{med}=0$ mag and \citet{hami02} find a $-$0.65 mag difference (both these values were obtained in the $V$ band which is close enough to our yellow band to allow a direct comparison, and both values include only radio-quiet quasars). The differences in the host luminosities computed by different groups are at least in part due to the modeling difficulties associated with subtracting the bright central source, but may also be related to the differences in the luminosities of quasars in different samples, if host and quasar luminosities are strongly correlated. 

\begin{figure}
\epsscale{0.9}
\plotone{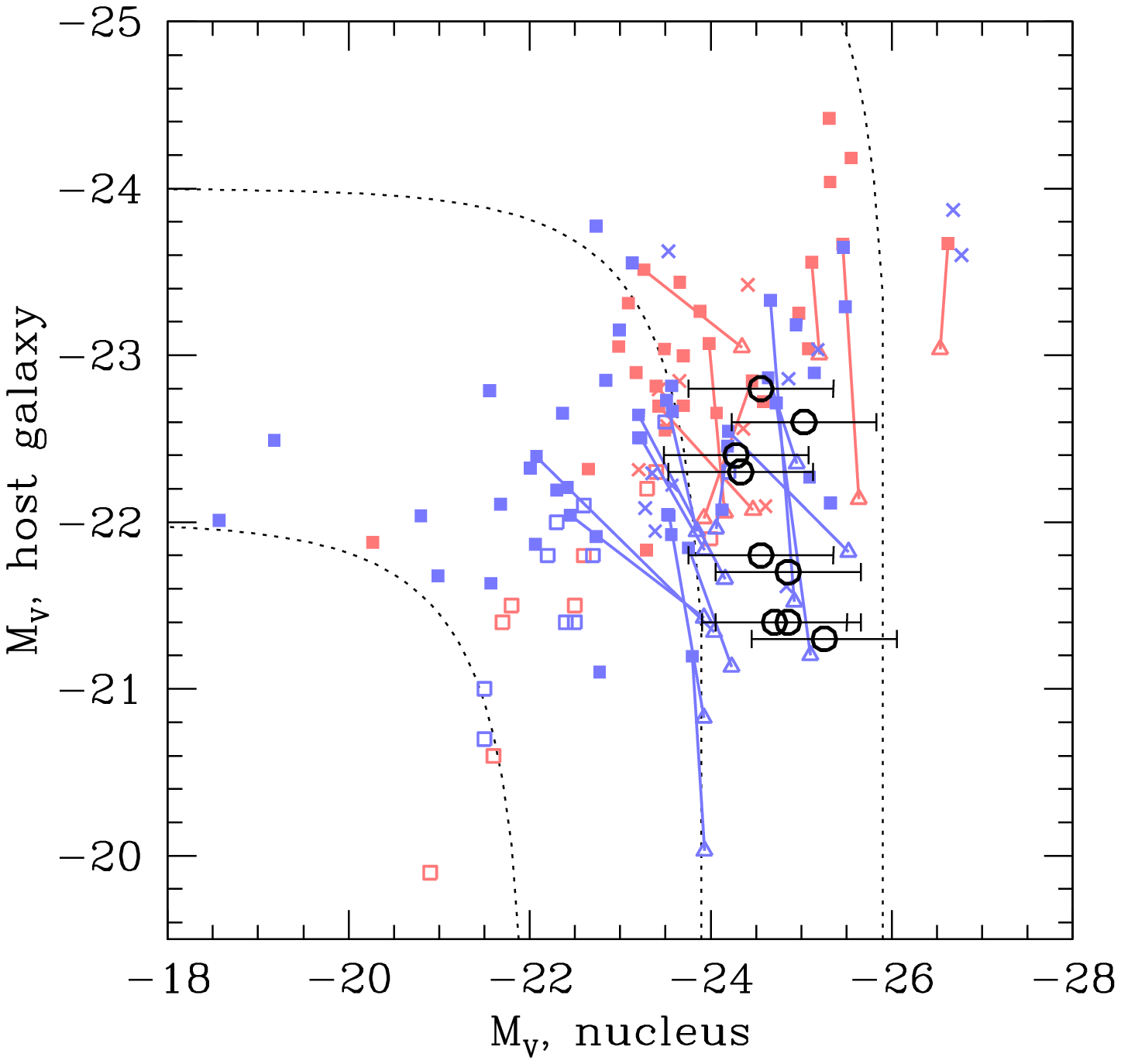}
\caption{Nuclear vs host luminosities of type II quasars (open circles with error bars) and of unobscured quasars (blue: radio-quiet, red: radio-loud). Different symbols correspond to objects from different studies (filled squares: \citealt{hami02}, open triangles: \citealt{bahc97}, crosses: \citealt{floy04}, open squares: \citealt{jahn04}). All objects from the sample by \citet{bahc97} were re-analyzed by \citet{hami02}; for a given object, the results of the two analyzes are connected with a solid line. All magnitudes from the literature were corrected to the $h=0.7$, $\Omega_m=0.3$, $\Omega_{\Lambda}=0.7$ cosmology. Dotted lines are the contours of constant total (nucleus+host) flux. Intrinsic nuclear luminosities of type II quasars are estimated using the [OIII]5007$-M_V$ correlation that exists for unobscured quasars (Paper I). }
\label{pic_qso}
\end{figure}

In Figure \ref{pic_qso} we show type II quasar nuclear and host luminosities, as compared with the same values for unobscured quasars from the literature. Because the quasars in our study are obscured in the optical, their intrinsic $M_V$ values are computed using the [OIII]5007 line luminosity (Paper I); the error bars reflect the 1$\sigma$ uncertainty due to the scatter in the [OIII]5007$-M_V$ correlation for unobscured quasars. Our yellow band is close to the $V$-band, and the K-corrections required to obtain $V$-band host magnitudes from the yellow-band magnitudes listed in Table \ref{tab:par} are less than 0.1 mag. For unobscured quasars, the PSF of the bright central source has to be included in the 2-D modeling, and the difference between host/nuclear luminosities obtained by \citet{bahc97} and \citet{hami02} for identical objects (solid lines in Figure \ref{pic_qso}) can be taken as a measure of the 2-D modeling uncertainties. Any 2-D fitting procedure based on $\chi^2$ minimization will be somewhat degenerate in separating the nuclear flux from the host flux, but will keep the total flux roughly constant. Indeed, most of the solid lines follow the contours of constant flux (examples are shown in dotted lines) rather closely. 

From Figure \ref{pic_qso}, we find that type II quasar hosts occupy a similar range of luminosities to the hosts of luminous quasars. We also find that the uncertainties in the positions of type II quasars on the nucleus/host diagram are comparable to those of unobscured quasars. The large uncertainties in the nuclear luminosities of type II quasars are counter-balanced by our $\sim 0.1$ mag photometry of the hosts.

Figure \ref{pic_lum_col}b shows the distribution of the colors of type II quasar hosts relative to those of the $M_*$ galaxies. Type II quasar hosts have UV$-$yellow colors similar to those of $M_*$ galaxies, but they are significantly bluer in the blue$-$yellow color, with a median color (relative to $M_*$ values) of $-0.4$ mag, reaching $-0.8$ mag for SDSS~J1039+6430. There is no evidence that the colors of the disk+bulge systems in our sample are different from the colors of the ellipticals. 

The values of the colors can be used to place constraints on the ages of the stellar population. For example, for the six objects at redshift around 0.4 (i.e., excluding group 2 objects) the colors can be directly compared with one another as the effective wavelengths of the observations are very similar in all six cases; the median colors for these objects are 0.8 mag (blue$-$yellow) and 1.8 mag (UV$-$yellow). To estimate the age corresponding to these colors, we use the stellar population synthesis models of \citet{bruz03} of a passively evolving instantaneous starburst and produce a series of spectra as a function of age. We then calculate the model colors using these model spectra and the effective wavelengths of our observations listed in Table 1. For a fixed age, of all the parameters probed by \citet{bruz03}, such as different evolution prescriptions, different libraries of UV stellar spectra and different initial mass functions, the one that affects the colors the most is metallicity. If the initial metallicity is assumed to be solar, then the median colors of our galaxies indicate an age between 3$\times 10^8$ and $10^9$ years. If the initial metallicity is 20\% solar, then the inferred age is around $10^9$ years. For SDSS~J1039+6430, the bluest object in the sample, the age estimates lie between $10^8$ and $3\times 10^8$ years. Our age estimates were obtained under the assumption that the entire galaxy underwent an episode of star formation, so if only a fraction of the mass of the galaxy was involved, then our estimated ages for the starburst should be regarded as upper limits. 

Our estimated ages suggest that A-type stars are an important contribution to the type II quasar hosts. They are best probed by our blue band observations which cover the spectral region near the Balmer break. In Paper II we demonstrated, using the high signal-to-noise ratio observations of the continuum emission, that the stellar light of SDSS~J1039+6430 is dominated by an A-star population.  

The effective radii listed in Table \ref{tab:par} (exponential scales in the case of the SDSS~J1106+0357 disk component) can give us a handle on the color gradients in the stellar populations. The accuracy of the measurement of the $a_1$ parameter is much worse than the accuracy of the model photometry (Section \ref{sec_2d}), and furthermore the effective radius is $\propto a_1^{-4}$ for a de Vaucouleurs profile, so we estimate that the accuracy of the measurement of the effective radii is no better than 20\%, limiting us to a qualitative description. Focussing only on the extended emission that is well-modeled by the de Vaucouleurs profile (or exponential in the case of SDSS~J1106+0357) and disregarding the central excesses of ambiguous origin, we notice that our sample presents a mix of galaxies that become bluer outward (e.g., SDSS~J1243$-$0232) and those with the color gradient of the opposite sign (e.g., SDSS~J1323$-$0159). 

\subsection{Scattered light}

Free electrons and dust particles in the interstellar medium of the host galaxy can scatter nuclear light. Distinguishing between these two possibilities can be rather difficult, as has been demonstrated by many authors (see Paper II for a detailed description and references), and it seems that both mechanisms operate in practice (e.g., \citealt{kish99, hine01}). For our sample, we find strong evidence for dust scattering in SDSS~J1039+6430 (Paper II and Section \ref{sec_1039}), but similar arguments are not conclusive when applied to other objects. Both for electron scattering and dust scattering the number density of particles required to account for the observed amount of scattered light can be computed from the geometric dimensions of the scattering regions, supplemented by a guess about scattering efficiency (Paper II). Interestingly, the required number density turns out to be $n_e \ga 1$ cm$^{-3}$ for electron scattering or $n_H \ga 1$ cm$^{-3}$ for dust scattering. Field ellipticals do not normally contain such large amounts of interstellar matter in a significant portion of their volume \citep{goud94}, and therefore these large values present a challenge in unifying elliptical hosts of quasars with their inactive counterparts.

In Table \ref{tab:par} we list lower limits on the ratio of the scattered to stellar flux, calculated by taking into account only those positive residuals that we are confident are due to scattered light, i.e., confirmed by their correct orientation relative to the polarization position angle. This procedure only works for scattering regions that are detected on kpc scales, so that their orientation relative to the nucleus can be used. In several objects (e.g., SDSS~J2358$-$0009) there is a central excess of light over our best-fitting stellar profile. Although this component is technically resolved in all such objects (i.e., it is inconsistent with a PSF), it is impossible to determine if this light is due to a circumnuclear starburst or to sub-kpc scattered regions that are made morphologically smooth by the effects of the PSF. In principle the two can be distinguished by comparing their colors, especially in the UV band. However, because central excesses are coincident with central dust obscuration and because the quality of stellar fits are often poorest near the center, this method cannot be satisfactorily used for our data.

We can also place strong lower limit on a scattered component based on the high observed polarization of the objects in our sample. The median observed polarization in the blue band is about 3\% (Table \ref{tab:pol}). Only the scattered component is polarized, and the polazed signal is diluted by the stellar light. The polarization of the scattered light in AGNs has rarely been observed to be more than about 30\%, and is of course limited to be $<$100\%. Therefore, to produce the observed 3\% polarization, at least 10\% of the observed flux should be due to scattered light. This argument demonstrates that our photometric procedure did not work satisfactorily in SDSS~J1106$+$0357 and SDSS~J2358$-$0009, where it failed to detect most of the scattered light. In SDSS~J1039+6430, the high ratio of the scattered to stellar light ($\ga 0.8$ in the blue) is consistent with the high observed level of polarization (10\% in the same band). 

Another method to estimate the contribution of scattered light is to include a power-law component in the stellar synthesis analysis of the AGN+host spectra. This method was employed, for example, by \citet{kauf03}, who set an upper limit of a few per cent on the scattered light in the spectra of type II AGN hosts in their sample. This finding is unsurprising given the much lower typical luminosities of the AGNs in their sample -- for a scattering efficiency of one per cent or so, only the most luminous AGNs are expected to have a significant contribution of scattered light to their total flux which is otherwise dominated by the host galaxy and narrow emission lines. 

We used a simplified version of this type of analysis in Paper II, where we attempted to represent the high signal-to-noise ratio MMT spectra of several type II quasars with a combination of a spectrum of an old elliptical galaxy, an A-star spectrum and a power-law component (representing the possible scattered light). In that paper, we found that 10-100\% of the broad-band flux was due to scattered light. The three objects that overlap between Paper II and this paper are SDSS~J1039+6430, SDSS~J1323$-$0159 and SDSS~J1413$-$0142; the scattered-to-stellar ratio in the yellow band published in Paper II for these objects was estimated to be 1.0, 1.0 and 0.7, correspondingly. The spectral analysis yielded much higher values than those listed in Table \ref{tab:par}, supporting our understanding that the aperture photometry of the positive residuals underestimates the true scattered flux (the aperture of the MMT spectra may have not enclosed all of the extended stellar flux of the host, but this effect probably cannot explain all of the difference in measured scattered-to-stellar ratios). Only for SDSS~J1039+6430 do the results of the two analyzes marginally agree, perhaps because of the lack of strong dust obscuration in the nucleus and the high quality of the 2-D fits. 

One of the most surprising results of our study is the great variety of shapes, sizes and opening angles of the scattering regions that we see in Figures \ref{pic_rgb1}-\ref{pic_rgb2}. Some of them appear as regular, edge-brightened cones (e.g., SDSS~J1323$-$0159), while others appear as clumpy irregular patches (e.g., SDSS~J1039+6430). Some scattering regions are biconical, whereas others are one-sided. The sample is divided almost equally between objects with large opening angles of scattering (30$-$60\degree) and those with jet-like, narrow regions with opening angles less than 10\degree. 

In seven galaxies, we detect kpc-scale dust extinction features. This finding is reminiscent of that by \citet{malk98} of the excess galactic dust in Seyfert 2 galaxies. The values of extinction that we find rarely exceed $A_V=1$ mag, much less than what is required to obscure the central AGNs. (If dust is concentrated in optically thick clouds, our measured spatially averaged extinction is not equivalent to the probable optical depth toward the AGN which may in this case be much larger.) More importantly, the scattering regions seen in our images have a well-defined conical or bi-conical structure with opening angles that are much smaller than 180\degree\ -- these features are hard to explain if obscuration occurs in thin lanes on kpc scales. Finally, in some objects (e.g., SDSS~J0920+4531) the observed areas of extinction are not centered on the nucleus, and in two objects (SDSS~J1243$-$0232 and SDSS~J2358$-$0009) we did not find any evidence for extinction in the host. Therefore, we conclude that this kpc-scale dust is not the primary agent of AGN obscuration, in accord with the standard unification picture in which the obscuration occurs on much smaller scales. While \citet{malk98} found an interesting excess of dustiness in Seyfert 2 galaxies compared to Seyfert 1 galaxies, it is currently impossible to address the presence of extinction in the central parts of luminous type I quasar hosts.

\begin{figure}
\epsscale{0.9}
\plotone{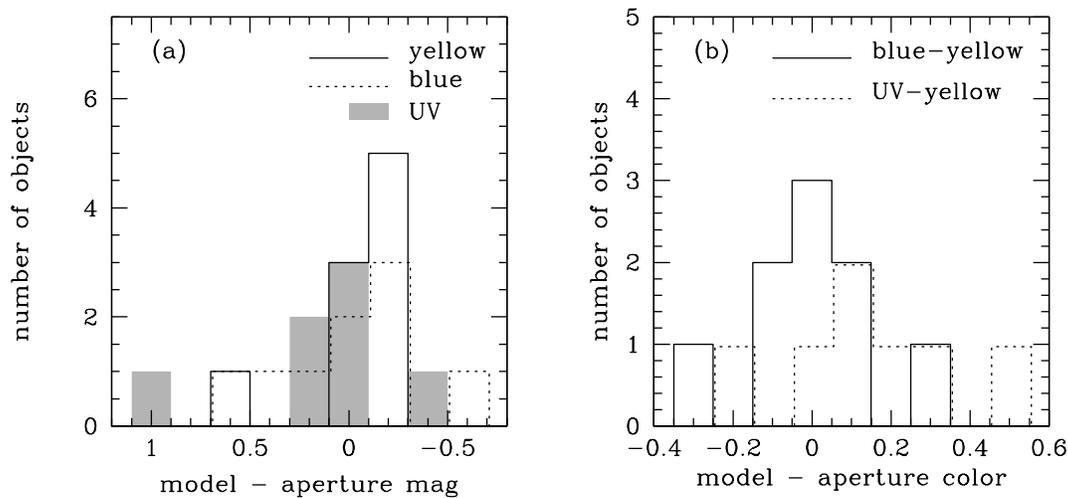}
\caption{(a) Distribution of model magnitudes of type II quasar hosts relative to the aperture magnitudes in the yellow band (solid; median value $-0.1$ mag), in the blue band (dotted; median value $-0.1$ mag) and in the UV band (grey histogram; median value 0.1 mag). Values $<0$ mean model magnitudes brighter than aperture magnitudes. (b) Distribution of model colors of type II quasar hosts relative to the aperture values: blue$-$yellow colors (solid; median value 0.0 mag) and UV$-$yellow colors (dotted; median value 0.1 mag). Relative colors $>0$ mean model colors are redder than aperture colors. }
\label{pic_model_aper}
\end{figure}

In Figure \ref{pic_model_aper} we show the difference between the aperture magnitudes (Table \ref{tab:aper}) and the model magnitudes obtained as a result of our 2-D fitting procedure (Table \ref{tab:par}). There are no apparent systematic differences between aperture and model photometry; in fact, the median model magnitudes and colors agree with those derived from aperture photometry within our claimed photometric uncertainty. It seems that, for a median galaxy, the effects of scattering and obscuration cancel each other out, somewhat by accident, as both effects become stronger toward shorter wavelengths. However, there is a large scatter in both luminosity differences and the color differences. If we were to use aperture photometry as a proxy for the luminosities and colors of the stellar component of type II quasar hosts (e.g., if detailed morphological information were not available), it would be accurate to no better than 0.3 mag.

\section{Conclusions}
\label{sec_conclusions}

In this paper, we analyze three-band ACS HST images of the host galaxies of nine optically-selected type II (obscured) quasars in combination with polarimetric data. Since optical radiation from type II quasars is heavily obscured, there is no contamination from the bright central source, and all host galaxies are easily detected. We further avoided the contribution from the narrow-line region by observing in filters that fall between strong emission lines. 

Type II quasars reside in elliptical hosts (6/9), in a disk+bulge host (1/9) and in disturbed disk+bulge hosts (2/9). One host can be described as undergoing a major merger, while up to three others show tidal debris. Type II quasar hosts are morphologically and dynamically similar to the hosts of ordinary radio-quiet quasars of similar luminosities \citep{bahc97, boyc98, boyc99, hami02} and of reddened quasars (\citealt{marb03}, but see also \citealt{hutc03} who argue for a much higher incidence of interacting systems among reddened quasars). The objects presented in this paper are estimated to have intrinsic luminosities in the range $-24>M_B>-26$ and are therefore of somewhat lower luminosity than the most extremely luminous unobscured radio-quiet quasars that seem to lie almost exclusively in massive elliptical hosts with small ellipticities \citep{kuku99, dunl03, floy04}.

The stellar light from type II quasar hosts is brighter and bluer than $M_*$ galaxies at the same redshift. The stellar population of the median galaxy in our sample has colors consistent with a $10^9$ year-old starburst, while the bluest galaxy may be as young as $10^8$ years old. Several surveys have found that AGN host galaxies are significantly bluer than inactive elliptical galaxies. For example, based on stellar population synthesis modeling, \citet{kauf03} estimated that many of the type II AGN host galaxies have undergone significant star formation within the last $(1-2)\times 10^9$ years. Our crude age estimates for the hosts of the high-luminosity analogs of the objects in their sample are in remarkable agreement with these values. Determining colors of the hosts of luminous type I AGNs is a very challenging task, as the already overwhelming nucleus becomes even more prominent toward shorter wavelengths. Studies of the hosts of medium-luminosity quasars ($-23>M_B>-24$) show that the elliptical hosts are as blue as the spiral hosts \citep{jahn04, sanc04}. We find that hosts of type II quasars are still blue at intrinsic nuclear luminosities of $-24>M_B>-26$. At yet higher nuclear luminosities, the colors and ages of quasar hosts are controversial. \citet{kirh99} found that even at these extreme luminosities the hosts of radio-loud quasars are significantly bluer than inactive galaxies of the same morphology and luminosity. Other authors \citep{kuku99, nola01} have argued that hosts of very luminous quasars, regardless of their radio properties, assembled around redshift 2.5 and have evolved passively ever since. In some of the same objects, \citet{cana00} found prominent young stellar populations and pointed out the importance of aperture effects in determining the ages of the stellar populations.

Scattered light contaminates the observed images and makes it difficult to study the properties of the stellar populations in the host galaxies. In extreme cases, we find that scattered light can be the dominant source of the extended flux, especially in the UV. We have considered various methods to address this contamination, such as morphological analysis (aided by polarimetric data) or stellar population synthesis analysis that includes a power-law template to account for a possible scattered component. For nearby well-resolved objects, imaging polarimetry can yield a robust determination of the extent and relative contribution of the scattered component (\citealt{cape95, kish99, kish02a, kish02b}; see also \citealt{tadh00} for imaging polarimetry of radio galaxies). At higher redshifts, this analysis becomes increasingly more difficult and is only possible for the largest scattering regions (\citealt{tran98}; see also review of polarimetric studies of radio galaxies by \citealt{tadh05}). 

The difficulties of isolating the scattered component are expected to be more severe in studies of hosts of unobscured quasars where the presence of a very bright blue nuclear source makes detailed morphological analysis difficult and greatly dilutes the polarization signal. Furthermore, since these objects are presumably seen more face-on with respect to the obscuration plane, there is no preferred scattering direction in the plane of the sky, resulting in a geometrical cancellation of the polarization from different directions. As a result, scattered flux can be significant without producing high polarization. In quasars with dust scattering (e.g., SDSS~J1039+6430), the forward nature of dust scattering can make the contamination by the scattered light even stronger in type I quasars than in type II quasars, because the typical scattering angles are smaller in face-on than in edge-on objects. 

The differences between the hosts of radio-loud and radio-quiet quasars have been addressed in many studies, but the origin and the nature of differences are still subject to debate. In this paper, we focused on radio-quiet type II quasars, comparing our results exclusively to other studies of radio-quiet quasars, unless explicitly stated otherwise. The radio-loud analogs of type II quasars from our sample, Narrow-Line Radio Galaxies, show many of the same features that we discuss in this paper, such as extended scattered light, significant contribution from the young stellar populations \citep{tadh02}, and a range of morphologies and dynamical states \citep{deko96}. A careful statistical comparison of large samples of radio-quiet and radio-loud type II AGNs is needed to address the spectacular difference in their radio output.

In the Introduction, we postulated that the hosts of type II AGNs form a representative sample of all AGN hosts. Statistical differences between type II and type I hosts have been claimed in the literature in the past, suggesting that some connection is present between the parsec-scale nuclear region and the kpc-scale host \citep{anto89}. Such comparison becomes increasingly more complicated at higher luminosities, but so far we have not seen any dramatic differences between the morphologies, dynamical states, luminosities and ages of the stellar populations of the hosts of type II vs type I AGNs. 

From spectropolarimetric observations presented in Paper II, we know that at least some, if not all, of our type II quasars are seen as broad-line AGNs along some directions. Therefore, our findings apply to at least some type I quasars. The question remains open as to whether all type I quasars are obscured along some lines of sight or whether there exists a population of truly `naked' type I quasars that are seen without circumnuclear dust obscuration along any direction. The calculation of the type I/type II ratio at high luminosities meets serious statistical difficulties, because there is at present no robust method of AGN selection that is not biased toward a specific type. 

\section*{Acknowledgments}

NLZ is supported by the {\it Spitzer} Space Telescope Fellowship provided by NASA through a contract issued by the Jet Propulsion Laboratory, California Institute of Technology. NLZ and MAS acknowledge support of NSF grant AST-0307409. PSS acknowledges support from Spitzer/JPL contract 1256424. Polarimetric studies at Steward Observatory are supported by the NSF through grant AST 03-06080. The authors thank the referee W.C. Keel for the useful comments. 

This paper is based on the observations made with the NASA/ESA Hubble Space Telescope (program 9905). Support for this work was provided in part by NASA through grant HST-GO-09905.01 from the Space Telescope Science Institute (STScI). STScI is operated by the Association of Universities for Research in Astronomy, Inc. under NASA contract NAS 5-26555. 

Funding for the SDSS and SDSS-II has been provided by the Alfred P. Sloan Foundation, the Participating Institutions, the National Science Foundation, the U.S. Department of Energy, the National Aeronautics and Space Administration, the Japanese Monbukagakusho, the Max Planck Society, and the Higher Education Funding Council for England. The SDSS Web Site is http://www.sdss.org/.

The SDSS is managed by the Astrophysical Research Consortium for the Participating Institutions. The Participating Institutions are the American Museum of Natural History, Astrophysical Institute Potsdam, University of Basel, Cambridge University, Case Western Reserve University, University of Chicago, Drexel University, Fermilab, the Institute for Advanced Study, the Japan Participation Group, Johns Hopkins University, the Joint Institute for Nuclear Astrophysics, the Kavli Institute for Particle Astrophysics and Cosmology, the Korean Scientist Group, the Chinese Academy of Sciences (LAMOST), Los Alamos National Laboratory, the Max-Planck-Institute for Astronomy (MPA), the Max-Planck-Institute for Astrophysics (MPIA), New Mexico State University, Ohio State University, University of Pittsburgh, University of Portsmouth, Princeton University, the United States Naval Observatory, and the University of Washington.

\begin{appendix}
\section{Broad-band polarimetry}
\label{sec_app}

Broad-band polarimetric measurements for 11 objects from Paper I are presented in Table \ref{tab:app}. Observations were conducted using a $V$ filter ($4800-6000$\AA). The measured values are $Q$, $U$ and $\sigma(P)$. The derived values are $\tan 2\theta=U/Q$ (where $\theta$ is in degrees East of North), $P=\sqrt{Q^2+U^2}$, $\sigma(\theta)=28.65^{\rm o} \times \sigma(P)/P$, and $q=\sqrt{P^2-\sigma(P)^2}$ (only for the objects with $P>\sigma(P)$). The polarization position angle is essentially undefined if $P<1.5\times \sigma(P)$ \citep{ward74}. The value $P$ is positive definite and therefore in the presence of measurement errors is an overestimate of the true polarization level $q$. The latter is estimated using the approximation by \citet{ward74}. The observational setup is described in Section \ref{sec_pol}. With the exception of SDSS~J0056+0032, polarization is detected in all objects at less than the $2\sigma$ level. 

\end{appendix}

\clearpage
\begin{deluxetable}{cccc|ccc|cc}
\rotate
\tabletypesize{\small}
\tablewidth{0pt}
\setlength{\tabcolsep}{0.03in}
\tablecaption{Summary of HST observations}
\tablehead{Object name & redshift, & log & filters & \multicolumn{3}{c|}{rest-frame eff. $\lambda$, \AA} & & \\
 (J2000) & $z$ & $L$[OIII]/$L_{\odot}$ & yellow, blue, UV & yellow & blue & UV & polarimetry & group}
\startdata
SDSS~J012341.47$+$004435.9 & 0.399 & 9.13 & FR914M, F550M, F435W & 5700 & 3989 & 3086 & SO2.3 & 1 \\
SDSS~J092014.11$+$453157.3 & 0.402 & 9.04 & FR914M, F550M, F435W & 5688 & 3981 & 3079 & MMT & 1\\
SDSS~J103951.49$+$643004.2 & 0.402 & 9.41 & FR914M, F550M, F435W & 5688 & 3981 & 3079 & MMT (Paper II) & 1\\
SDSS~J110621.96$+$035747.1 & 0.242 & 9.13 & FR647M, F550M, FR459M & 5804 & 4494 & 3124 & MMT & 2 \\
SDSS~J124337.34$-$023200.2 & 0.281 & 9.02 & FR647M, F550M, FR459M & 5789 & 4357 & 3118 & & 2 \\
SDSS~J130128.76$-$005804.3 & 0.246 & 9.25 & FR647M, F550M, FR459M & 5786 & 4479 & 3114 & & 2 \\
SDSS~J132323.33$-$015941.9 & 0.350 & 9.19 & F775W, F550M, F435W & 5699 & 4134 & 3198 & MMT (Paper II) & \nodata \\
SDSS~J141315.31$-$014221.0 & 0.380 & 9.25 & FR914M, F550M, F435W & 5779 & 4044 & 3129 & MMT (Paper II) & 1 \\
SDSS~J235818.87$-$000919.5 & 0.402 & 9.32 & FR914M, F550M, F435W & 5688 & 3981 & 3079 & SO2.3 & 1
\enddata

\tablecomments{J2000 coordinates, redshifts and [OIII]5007 luminosities are from the SDSS data (Paper I). For five objects spectropolarimetry data from MMT are available; for three of these the data were published in Paper II, and for the remaining two the data are presented in this paper. For two additional objects broad-band polarimetry obtained using the Steward 2.3m telescope (SO2.3) is available. In the last column, we grouped the objects for display purposes based on similarity of the filters used in the observations. Objects from the same group have the same parameters of the color-composite \citep{lupt04}, and therefore can be visually compared with one another (Figures \ref{pic_rgb1}-\ref{pic_rgb2}). }
\label{tab:ids}
\end{deluxetable}

\clearpage
\begin{deluxetable}{ccccc}
\rotate
\tablewidth{0pt}
\setlength{\tabcolsep}{0.03in}
\tablecaption{Summary of polarization properties}
\tablehead{Object & UV & blue & yellow & mean $\theta$ \\
name & polarization & polarization & polarization & deg. E of N  }
\startdata
SDSS~J0123$+$0044 & \nodata & 2.7$\pm$2.2 & \nodata & 8$\pm$ 25\\
SDSS~J0920$+$4531 & 4.7$\pm$0.9 &  2.5$\pm$0.4 &  0.7$\pm$0.6 & 171$\pm$2\\
SDSS~J1039$+$6430 & 16.6$\pm$0.3 & 10.2$\pm$0.2 & 8.0$\pm$0.3 & 109$\pm$1 \\
SDSS~J1106$+$0357 & 3.1$\pm$0.5 &  1.0$\pm$0.3 &  0.1$\pm$0.2 & 146$\pm$3 \\
SDSS~J1323$-$0159 & 5.0$\pm$1.0 &  3.7$\pm$0.5 &  2.8$\pm$0.5 & 104$\pm$3 \\
SDSS~J1413$-$0142 & 4.1$\pm$1.0 &  2.9$\pm$0.4 &  1.2$\pm$0.6 & 146$\pm$8 \\
SDSS~J2358$-$0009 & \nodata & 2.8$\pm$1.8 & \nodata & 154$\pm$15\\
\enddata

\tablecomments{`UV', `blue' and `yellow' polarizations (given in \%) are calculated as close to the ACS filter coverage as possible. We list the mean polarization position angle $\theta$ and its 1$\sigma$ error (calculated over the entire available polarimetric wavelength coverage). }
\label{tab:pol}
\end{deluxetable}

\clearpage
\begin{deluxetable}{cccc}
\rotate
\tablewidth{0pt}
\tablecaption{Aperture photometry of type II quasars}
\tablehead{Object & \multicolumn{3}{c}{absolute AB magnitudes} \\
name & yellow & blue & UV }
\startdata
SDSS J0123$+$0044 & -21.8 & -20.7 & -19.9 \\ 
SDSS J0920$+$4531 & -22.4 & -21.6 & -20.7 \\ 
SDSS J1039$+$6430 & -22.0 & -21.6 & -21.0 \\ 
SDSS J1106$+$0357 & -22.7 & -22.0 & -19.6 \\ 
SDSS J1243$-$0232 & -22.6 & -21.9 & -19.9 \\ 
SDSS J1301$-$0058 & -21.7 & -21.1 & -19.3 \\ 
SDSS J1323$-$0159 & -21.4 & -20.7 & -19.8 \\ 
SDSS J1413$-$0142 & -21.6 & -20.9 & -19.9 \\ 
SDSS J2358$-$0009 & -22.5 & -21.8 & -20.9 \\ 
\enddata

\tablecomments{Magnitudes are given at effective wavelengths listed in Table \ref{tab:ids}.}
\label{tab:aper}
\end{deluxetable}

\clearpage
\begin{deluxetable}{ccc|ccc|cccccc|c}
\rotate
\tabletypesize{\footnotesize}
\tablewidth{0pt}
\setlength{\tabcolsep}{0.1in}
\tablecaption{Structural parameters and luminosities of type II quasar host galaxies}
\tablehead{Object & S\'ersic & & \multicolumn{3}{c|}{$R_d$ or $R_e$, kpc}& \multicolumn{6}{c|}{model AB magnitudes of the stellar component} & scattered-to-\\
name & index & ell. & yellow & blue & UV & yellow & (yellow$*$) & blue & (blue$*$) & UV & (UV$*$) & stellar}
\startdata
SDSS~J0123$+$0044 & 4 & 0.45 & 3.7 & 6.0 & 4.9 & -21.9 & (-21.7) & -20.8 & (-20.6) & -19.7 & (-19.9) & $>0.1$\\
SDSS~J0920$+$4531 & 4+1 & 0.6 & \nodata & \nodata & \nodata & -22.4 & (-21.7) & -21.7 & (-20.6) & -20.7 & (-19.9) & \\
SDSS~J1039$+$6430 & 4 & 0.1 & 1.4 & 2.3 & 1.9 & -21.4 & (-21.7) & -21.1 & (-20.6) & -20.1 & (-19.9) & $\ge$ 0.8\\
SDSS~J1106$+$0357 & 1 & 0.45 & 11.0 & 14.3 & \nodata & -22.9 & (-21.5) & -22.5 & (-20.6) & $<$-19.2 & (-19.2) & $>$0.01\\
                  & 4 & 0.4 & 5.5 & 3.8 & 1.6 & total & & total & & total & & \\
SDSS~J1243$-$0232 & 4 & 0.1 & 6.5 & 7.0 & 8.1 & -22.5 & (-21.6) & -21.7 & (-20.6) & -19.9 & (-19.4) & \\
SDSS~J1301$-$0058 & 4 & 0.2 & 10.3 & 14.4 & \nodata & -21.8 & (-21.5) & -21.3 & (-20.6) & \nodata & (-19.2) & \\
companion & 4+1 & 0.5 &  \nodata &  \nodata & \nodata & \nodata & & -19.2 & & \nodata & & \\
SDSS~J1323$-$0159 & 4 & 0.35 & 1.9$-$2.6 & 1.2$-$3.2 & 0.6$-$1.8 & -21.5 & (-21.7) & -20.9 & (-20.6) & -20.1 & (-19.7) & $>$0.08 \\
SDSS~J1413$-$0142 & 4+1 & 0.45 & \nodata & \nodata & \nodata & -21.5 & (-21.7) & -20.5 & (-20.6) & -19.8 & (-19.8) & $>$0.09 \\
SDSS~J2358$-$0009 & 4 & 0 & 8.6 & 11.1 & \nodata & -22.7 & (-21.7) & -21.9 & (-20.6) & -20.8 & (-19.9) & $>$0.005\\
companion & 4+1 & 0 & \nodata & \nodata & \nodata & -21.6 & & -20.7 & & -19.0 & & \\
\enddata

\tablecomments{Parameters of the best 2-D model fits (de Vaucouleurs if the S\'ersic index is 4, exponential if it is 1). The column `ell.' lists our best-fit ellipticities in the blue band. Typical uncertainties in the characteristic radius are 20\% or more. Absolute model magnitudes are given at the effective rest-frame wavelengths listed in Table \ref{tab:ids}. The values with the $*$ symbol are absolute magnitudes of the $M_*$ galaxies at the same redshift and the same effective wavelength (Section \ref{sec_disc1}). Model magnitude uncertainties are 0.1 mag except for SDSS~J0920+4531, SDSS~J1323$-$0159 (UV) and SDSS~J1413$-$0142. In these cases, we estimate the uncertainties to be about 0.2-0.3 mag. In SDSS~1301$-$0058, only the central excess emission of unknown nature was detected in the UV band, and in SDSS~J1106+0357 only the bulge component was detected in the UV. In the last column, we list lower limits on the ratio of the scattered to stellar fluxes in the blue band, based on the aperture photometry of the positive residuals confidently identified as scattered regions. In SDSS~J0920+4531, bright positive residuals were detected, but their orientation does not agree with the polarization position angle. Polarization data were not available for SDSS~J1301$-$0058 and SDSS~J1243$-$0232, so none of the residuals could be identified as scattered light. }
\label{tab:par}
\end{deluxetable}

\clearpage
\begin{deluxetable}{cccccccccc}
\setlength{\tabcolsep}{0.05in}
\tablecaption{Appendix: polarization measurements of 11 type II AGNs from Paper I}
\tablehead{J2000 & & log & $Q$ & $U$ & $P$ & $q$ & $\sigma(P)$ & $\theta$ & $\sigma(\theta)$ \\
coordinates & $z$ & $L$[OIII]/$L_{\odot}$ & \% & \% & \% & \% & \% & deg. & deg. }
\startdata
SDSS J002531.46$-$104022.2 & 0.303 & 8.73 & -0.95 & -0.80 &  1.24 & 1.07 &  0.62 & 110 & 14 \\ 
SDSS J002827.78$-$004218.8 & 0.418 & 8.75 & -2.17 & -1.67 &  2.74 & 1.47 &  2.31 &  &  \\ 
SDSS J005009.81$-$003900.6 & 0.729 & 9.94 &  2.93 &  1.10 &  3.13 & 2.45 &  1.94 & 10 & 18 \\ 
SDSS J005621.72$+$003235.8 & 0.484 & 9.45 &  0.72 & -8.66 &  8.69 & 8.31 &  2.56 & 137 & 8 \\ 
SDSS J012341.47$+$004435.9 & 0.399 & 9.13 &  3.33 &  0.97 &  3.47 & 2.67 &  2.21 & 8 & 18 \\ 
SDSS J014237.49$+$144117.9 & 0.389 & 8.76 & -0.64 & -0.16 &  0.66 &  &  0.76 &  &  \\ 
SDSS J015911.66$+$143922.5 & 0.319 & 8.56 & -0.02 & -2.62 &  2.62 &  &  3.64 &  &  \\ 
SDSS J021758.19$-$001302.7 & 0.344 & 8.75 &  0.53 & -1.43 &  1.53 & 1.18 &  0.97 & 145 & 18 \\ 
SDSS J225102.40$-$000459.9 & 0.550 & 9.13 & -2.87 & -1.65 &  3.31 & 2.69 &  1.93 & 105 & 17 \\ 
SDSS J235818.87$-$000919.5 & 0.402 & 9.32 &  2.01 & -2.61 &  3.30 & 2.79 &  1.75 & 154 & 15 \\ 
SDSS J235831.16$-$002226.5 & 0.628 & 9.96 &  4.75 &  1.19 &  4.90 & 4.23 &  2.47 & 7 & 14 
\enddata
\tablecomments{Broad-band polarization measurements for 11 objects from Paper I. Coordinates, redshifts and [OIII] luminosities (quoted as $\log L$[OIII]/$L_{\odot}$) are from Paper I. Polarization values are quoted as $P$ (uncorrected for statistical bias) and $q$ (corrected for statistical bias). $\theta$ is the polarization position angle, measured East of North. }
\label{tab:app}
\end{deluxetable}

\end{document}